\documentclass{article}

\usepackage{amsmath}
\usepackage{amssymb}
\usepackage{graphicx}
\usepackage{authblk}
\usepackage{hyperref}
\usepackage{setspace}
\usepackage{fullpage}

\title{Crystallizing spacetime: a fundamentally classical framework for quantum gravity}
\author{Filip Strubbe\footnote{\url{filip.strubbe@ugent.be}}}

\affil[]{Electronics and Information Systems, Ghent University, Tech Lane Ghent Science Park-Campus A~126, 9052~Ghent, Belgium}

\begin{document}
\maketitle

\begin{abstract}
Conventional approaches to quantum gravity regard quantum principles, such as nonlocality and superposition, as fundamental properties of nature and therefore argue that gravity must also be quantized. In contrast, this work introduces a theory of crystallizing spacetime, which offers an alternative perspective: that both gravitational and quantum mechanical observations can be explained within a fundamentally classical framework operating beyond traditional spacetime. The theory proposes a spacetime relaxation mechanism wherein a dynamically evolving four-dimensional spacetime, populated by dynamic worldlines, relaxes as a function of the parameter $\tau$ into a standard spacetime consistent with general relativity. Simulations in the weak-gravity limit illustrate this process of spacetime and worldline relaxation. Additionally, models are developed showing that two hallmark quantum phenomena —nonlocality in an EPR experiment and double-slit interference of a massive particle— can be reproduced in this fundamentally classical framework by implementing Costa de Beauregard's concept of zigzag action along worldlines. The resulting crystallizing spacetime framework not only resolves the measurement problem but also provides a compelling basis for a unified theory of matter and gravity. Since this framework is grounded in concepts analogous to realism, locality, and determinism at its foundations, it brings Einstein's long-sought intuitive worldview within reach.
\end{abstract}

\newpage

\section{Introduction}
One of the biggest challenges in physics is explaining microscopic quantum mechanical observations as well as macroscopic gravitational observations consistent with general relativity in a unified framework. The prevailing view is that matter is fundamentally quantum mechanical and that gravity must consequently also be quantized. This notion traces back to arguments presented during the Chapel Hill Conference in 1957~\cite{Rickles2011,DiMauro2021a,Marletto2017b,Salcedo1996,Terno2006}. The arguments for quantizing gravity usually highlight conflicts with core principles of quantum mechanics, such as superposition and uncertainty, when attempting to use classical gravity to measure massive particles. As Feynman famously remarked: ``We're in trouble if we believe in quantum mechanics but don't quantize gravity''\cite{Zeh2011}. An alternative view suggests coupling quantum matter to classical gravity (see~\cite{Boughn2009,Carlip2008, Rosenfeld1963}). This approach, however, requires sacrificing reversibility~\cite{Galley2023,Oppenheim2023}. Oppenheim \textit{et al.} demonstrated an example of such a quantum-classical hybrid, termed post-quantum classical gravity~\cite{Oppenheim2023,Oppenheim2023a}. Despite their differences, both the quantized gravity approach and the quantum-classical hybrid approach share a foundational commitment to quantum principles like superposition, nonlocality, and uncertainty. In other words, they assume irreducibly quantum mechanical matter, that is fully non-classical and cannot be broken down into classical explanations~\cite{Galley2023}. These deeply counterintuitive quantum principles —involving abstract spaces, complex numbers, nonlocality, and superposition— lie at the heart of long-standing problems with unifying quantum mechanics with gravity, particularly with quantum measurement~\cite{Zurek2003,Bassi2013,Schlosshauer2005} and with time~\cite{Kiefer2013,Isham1993,Carlip2001, Zych2019}. 

This raises the question: How firmly should we adhere to the principles of quantum mechanics? On the one hand, the strong belief in quantum mechanics is supported by its overwhelming empirical success and by no-go theorems like Bell's theorem~\cite{Bell1964} that rule out traditional classical interpretations of quantum mechanics. More specifically, the violation of Bell inequalities, verified in EPR-type experiments~\cite{Groeblacher2007,Salart2008,Gisin2011,Drezet2019,Genovese2019,Ma2013,Giustina2015}, forces us to abandon at least one of the assumptions of locality, realism, or statistical independence, none of which can be dismissed without compromising traditional classical characteristics~\cite{Stanford, Valdenebro2002}. Consequently, attempts to explain quantum observations in a traditional classical way within standard spacetime, like Couder's bouncing oil drops~\cite{Couder2005, Couder2006}, must inevitably fail to replicate key quantum features, particularly quantum nonlocality~\cite{Andersen2015}. On the other hand, it is crucial to recognize that conclusions regarding what quantum mechanics tells us about reality typically implicitly presuppose a context of standard spacetime. This tacit assumption of standard spacetime is made in the Bohr-Einstein debates~\cite{Einstein1935,Bohr1949,Jammer1974}, in Bell's theorem~\cite{Bell1964,Wharton2020}, and often in contemporary discourse. Thus, while quantum mechanical observations may seem counterintuitive from a context of traditional spacetime, a more intuitive understanding may be obtained by abandoning the assumption of standard spacetime and exploring potential alternatives to quantum mechanics in such modified spacetimes. The weirdness of quantum mechanics may then be only an emergent phenomenon rather than an intrinsic feature of nature. This proposal aligns with suggestions from prominent physicists—including ’t Hooft, Weinberg, Smolin and Penrose—who have argued that both spacetime and quantum theory may need a radical reformulation (see~\cite{Strubbe2023}). Similar ideas of emergence have been explored in various contexts, such as Bohmian mechanics~\cite{Bohm1952}, 't Hooft's cellular automaton theory~\cite{Hooft2016}, and several other models~\cite{Requardt1996,Hooft1988,Hartle2006, Surya2019,Adler2012, Verlinde2011,Jacobson1995, Elze2017,Groessing2014, Smolin2012,Shor2023,Torrome2017}. It is further supported by the fact that, prior to any theoretical interpretation, outcomes of quantum measurements seem to be classical—this is the essence of the measurement problem and of the quantum-to-classical transition~\cite{Zurek2003,Bassi2013,Schlosshauer2005}. Namely, each measurement appears to produce a classical pointer state, associated to a classical measurement device (this can be click, a position of a dial, a light that switches on, etc...). In other words, quantum experiments do not directly reveal quantum properties like superposition, spin, or entanglement—these quantum properties are only inferred indirectly from analyzing the correlations and statistics of these measurement outcomes. Similarly, gravitational observations are successfully described by classical theory—namely, general relativity—again without requiring quantum concepts. This supports the idea that core concepts of quantum mechanics, like imaginary numbers, Hilbert spaces, superposition, and nonlocality, may not reflect intrinsic properties of nature, but rather features of a particular mathematical formalism.

Guided by the above-mentioned idea that nature may be based on intuitive principles, this work proposes a paradigm shift by turning traditional spacetime into a dynamically crystallizing spacetime, and by replacing the standard quantum formalism by a formalism based on interacting worldlines. The resulting theory, referred to as crystallizing spacetime, explains how macroscopic gravitational observations consistent with general relativity, as well as microscopic observations in agreement with quantum mechanics, can emerge from a unified, intuitive theory operating beyond standard spacetime. While the resulting framework is not classical in the conventional sense within standard spacetime, it can be regarded as fundamentally classical in a context beyond standard spacetime. Here, "fundamentally classical" refers to cherished concepts similar to locality, realism, and determinism, but operating at a deeper level of reality beyond ordinary spacetime. Hence, crystallizing spacetime attempts to lay the foundation for a theory of quantum gravity that is—surprisingly—fundamentally classical.

This is achieved, firstly, by introducing a spacetime relaxation mechanism describing the dynamic evolution of a four-dimensional spacetime as a function of the parameter $\tau$. With this mechanism, classical general relativity naturally emerges in the limit for $\tau \rightarrow \infty$, providing an elegant explanation for macroscopic gravitational observations. And, secondly, this is achieved by describing particles by dynamic worldlines that crystallize, transport information, and relax into geodesics as a function of $\tau$. The present work thus pursues a third strategy in addition to the above-mentioned approaches: making both gravity and quantum matter fundamentally classical—the complete opposite of quantizing gravity. It challenges the prevailing notion, based on the peculiar features of quantum mechanics, that an intuitive understanding of reality is unattainable or that gravity must inevitably be quantized. The presented framework further develops concepts introduced by Broad~\cite{Broad} and Ellis \textit{et al.}~\cite{Ellis2010}. Additionally, in earlier works of Strubbe~\cite{Strubbe2022,Strubbe2023} the idea of crystallizing spacetime has already been used to reproduce key quantum phenomena, such as double-slit interference, Mach-Zehnder interferometry, and EPR nonlocality in a fundamentally classical manner, but without gravity. In the current work, a key improvement is the implementation of gravity.

To illustrate the key mechanisms of crystallizing spacetime, simulations of spacetime and geodesic relaxation are performed under a weak-gravity approximation. Additionally, theoretical models are developed to reproduce two essential quantum phenomena in a fundamentally classical way, including gravitational effects: nonlocality in an EPR experiment and double-slit interference of a massive particle. The former emphasizes that understanding nonlocality does not require invoking mysterious action-at-a-distance. Instead, it shows how influences can propagate along worldlines as a function of $\tau$, in line with Costa de Beauregard's concept of zigzag action~\cite{Costa,Price2015}, and how quantum collapse may arise as a physical mechanism occurring "outside of spacetime", as suggested by Bancal \textit{et al.}~\cite{Bancal2012}. The latter illustrates that a particle in a superposition can be interpreted as a bundle of dynamically interacting worldlines evolving as a function of $\tau$, where only a single worldline carrying energy-momentum determines the outcome of the measurement and generates a gravitational effect. Finally, the core properties of the crystallizing spacetime framework are discussed, along with its connection to other theoretical approaches and its distinct predictions for future experimental tests of quantum gravity.

\section{Crystallizing spacetime}

This section introduces the theoretical basis of the crystallizing spacetime framework. As a note to the reader: several core equations governing the crystallizing spacetime framework are postulated upfront. In subsequent sections these equations will be demonstrated to have the desired physical behavior, through analytical arguments and illustrative examples.

\subsection{Geometric framework}

\begin{figure}[t!]
\centering
\includegraphics[width=16.0cm]{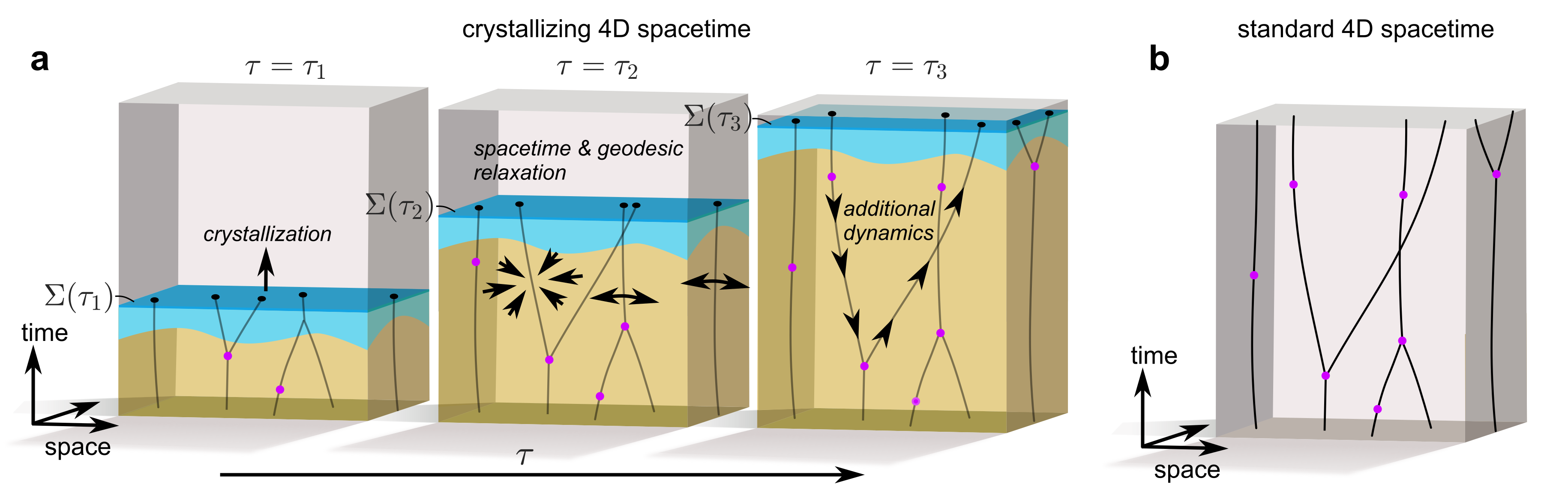}
\caption{Schematic illustration of crystallizing spacetime. \textbf{(a)} Dynamic evolution of a four-dimensional spacetime containing a few worldlines at three increasing values of the evolution parameter $\tau$. Key mechanisms—including spacetime and geodesic relaxation, crystallization, and additional dynamics relevant for reproducing quantum phenomena—are indicated. The crystallization hypersurface $\Sigma(\tau)$ is represented by the blue surface. Dynamics as a function of $\tau$ are largely restricted to a finite region (light-blue) in the past of $\Sigma(\tau)$, while further in the past dynamics gradually settle. In the limit of $\tau \rightarrow \infty$, a traditional four-dimensional spacetime emerges sufficiently in the past of $\Sigma(\tau)$, consistent with general relativity. Measurement outcomes by standard observers are represented by events (magenta dots). \textbf{(b)} From the perspective of a standard observer, measurement outcomes (magenta dots) are consistent with a conventional four-dimensional spacetime. Since these outcomes are acquired sequentially as a function of $\tau$ with increasing time coordinate values, this is consistent with a classical four-dimensional world view and with the flow of time.}\label{FIG0}
\end{figure}

Crystallizing spacetime is conceptualized as a fixed four-dimensional Lorentzian manifold $\mathcal{M}$ endowed with a metric $g_{\mu \nu}(\tau)$ and populated by physical entities with coordinates $x^\mu(\tau)$, that depend on an external evolution parameter $\tau$. Here, coordinates $x^\mu$ for $\mu = 0,1,2,3$ correspond respectively to $t$, $x$, $y$, $z$ with associated metric signature $(-,+,+,+)$. Such a framework with an evolving metric on a fixed manifold is well-established in the context of Ricci flow~\cite{Elliott2012,Topping2006,Perelman2003}, typically applied to compact Euclidean or Riemannian manifolds. In the present work, however, this approach is applied to a pseudo-Riemannian manifold. A schematic illustration of crystallizing spacetime and a comparison to traditional spacetime is given in Fig.~\ref{FIG0}. The parameter $\tau$, with unit $s^*$, is an external evolution parameter that should be interpreted analogously to the flow parameter in Ricci flow or the time parameter $t$ in classical mechanics: it is a monotonically increasing parameter which governs causality and drives the evolution of physical configurations on the manifold.

\subsection{Spacetime relaxation}
Next, within the framework of crystallizing spacetime, gravity is incorporated through the following spacetime relaxation mechanism:
\begin{equation}
 \kappa_{str}\frac{\partial g_{\mu \nu}(\tau)}{\partial \tau}= -\big( G_{\mu\nu}(\tau)-\frac{8\pi G}{c^4}T_{\mu\nu}(\tau) \big).
 \label{MasterEq}
\end{equation}
The form of Eq.~\eqref{MasterEq} is postulated here, bearing a clear resemblance to the Ricci flow equation~\cite{Dias2016,Ni2019}. Below and in the following sections we will evaluate how this equation leads to the desired results. At the right-hand side of Eq.~\eqref{MasterEq}, one recognizes familiar terms from the Einstein field equations. However, a $\tau$-dependency has been introduced for each component of the Einstein tensor $G_{\mu\nu}(\tau)=R_{\mu\nu}(\tau)-\frac{1}{2}R(\tau)g_{\mu\nu}(\tau)$ and energy-stress tensor $T_{\mu\nu}(\tau)$, where the Ricci tensor $R_{\mu\nu}(\tau)$ and Ricci scalar $R(\tau)$ in turn depend on the metric $g_{\mu\nu}(\tau)$. The parameter $\tau$, with unit $s^*$, is an external evolution parameter governing causality, while time $t$ represents merely the coordinate of a physical dimension. This evolution parameter is introduced as an additional degree of freedom needed to explain quantum observations in a fundamentally classical way, and to explain the observed flow of time, as will be discussed in more detail further below. The left-hand side of Eq.~\eqref{MasterEq} drives the relaxation of the metric $g_{\mu\nu}(\tau)$ towards equilibrium, at a rate determined by the scalar $\kappa_{str}$. Similar as for Ricci flow~\cite{Dias2016,Ni2019}, Eq.~\eqref{MasterEq} is fully covariant. The notation in Eq.~\eqref{MasterEq} assumes that the same coordinate system is used at each value of $\tau$, even though these coordinates may in principle be chosen arbitrarily. In the limit for $\tau \rightarrow \infty$, and assuming that the metric and energy-stress tensor achieve an equilibrium in which $\partial g_{\mu \nu}(\tau)/\partial \tau=0$ and $\partial T_{\mu \nu}(\tau)/\partial \tau = 0$ (as will be motivated further below), Eq.~\eqref{MasterEq} reduces to the Einstein field equations:
\begin{equation}
     G_{\mu\nu}=\frac{8\pi G}{c^4}T_{\mu\nu}.
     \label{EFEs}
\end{equation}
The spacetime relaxation mechanism thus ensures that a standard spacetime of general relativity emerges in equilibrium, consistent with macroscopic gravitational observations. Whereas other relaxation schemes, for example for solving differential equations~\cite{Rueter2018,Anderson2007,Christodoulou1979,Ishii2019,Lehner2001,Zhang2024}, in Ricci flow~\cite{Dias2016,Ni2019}, or in Parisi-Wu theory~\cite{ParisiWu}, are usually just mathematical tools to obtain a physical solution in equilibrium, here spacetime relaxation is considered to be an actual physical mechanism.

\subsection{Crystallization and geodesic relaxation}\label{SectionCryst}

To turn the spacetime relaxation mechanism into a framework consistent with observations, a few additions must be made. Firstly, particles are modeled exclusively in terms of worldlines that are crystallizing and relaxing together with spacetime as a function of $\tau$. Here, crystallization refers to the concept that there is a past region of spacetime where worldlines have been formed and relax towards equilibrium, while in a future region worldlines are not yet formed, with a boundary between these two regions shifting gradually toward the future time direction (see Fig.~\ref{FIG0}a). Similar as in Refs.~\cite{Strubbe2022,Strubbe2023}, we define the crystallization hypersurface $\Sigma(\tau)$ forming this boundary by:
\begin{equation}\label{crystallization}
    t_{cryst}(\tau)=\beta \tau,
\end{equation}
where $t$ is the time coordinate corresponding to the proper time of a distant observer at rest with respect to the expansion of the universe, and where $\beta$ is a positive scalar. 

Secondly, worldlines are governed by a geodesic relaxation mechanism:
\begin{equation}
    \kappa_{geo}\frac{\partial}{\partial \tau}(\frac{\partial^2 x^{\mu}}{\partial \lambda^2}) = - \Big(\frac{\partial^2 x^{\mu}}{\partial \lambda^2} + \Gamma^{\mu}_{\rho\sigma} \frac{\partial x^{\rho}}{\partial \lambda}\frac{\partial x^{\sigma}}{\partial \lambda}\Big).
    \label{geodesicrelaxation}
\end{equation}
Here, $x^{\mu}$ are $\tau$-dependent coordinates of the worldline, $\lambda$ is an affine parameter along the worldline, and the worldline relaxation rate is set by $\kappa_{geo}$. The left-hand side of Eq.~\eqref{geodesicrelaxation} is chosen to make the expression independent of the chosen affine parameter $\lambda$. As desired, in the equilibrium where $\frac{\partial}{\partial \tau}(\frac{\partial^2 x^{\mu}}{\partial \lambda^2})=0$, we find that such worldlines satisfy the geodesic equation:
\begin{equation}
    \frac{\partial^2 x^{\mu}}{\partial \lambda^2} + \Gamma^{\mu}_{\rho\sigma} \frac{\partial x^{\rho}}{\partial \lambda}\frac{\partial x^{\sigma}}{\partial \lambda}=0.
\end{equation}
At each value of $\tau$, an energy-stress tensor $T_{\mu\nu}(\tau)$, associated to given particle worldlines and their energy and momentum, is determined as usual in general relativity. Assuming that all worldlines ultimately reach equilibrium, also $T_{\mu\nu}(\tau)$ becomes fixed for $\tau\rightarrow\infty$. According to Eq.~\eqref{EFEs}, this means that the principle of energy-momentum conservation is satisfied in equilibrium. In the future region with respect to $\Sigma(\tau)$ a zero energy-stress tensor is imposed. 

\subsection{Additional evolution laws}
The spacetime relaxation, crystallization, and geodesic relaxation mechanisms, governed by Eqs.~\eqref{MasterEq},~\eqref{crystallization} and \eqref{geodesicrelaxation}, form the basis of the crystallizing spacetime framework. However, additional equations are required to describe the $\tau$-dependent dynamics of physical entities, worldlines, and influences propagating along worldlines. Additionally, elementary particles such as photons or electrons may need to be represented as bundles of interacting worldlines, each governed by the proposed evolution laws. These laws are therefore essential for reproducing the observed behavior of both classical and quantum mechanics within this framework, and will be examined in more detail in Sections~\ref{EPR} and~\ref{NeutronDS}.

\subsection{Observation and the emergence of traditional spacetime}\label{observation}

Next, we define observers and their observations within the crystallizing spacetime framework, and indicate how these observations agree as desired with a traditional classical world. Firstly, observers are assumed to be composed of the same worldlines that constitute elementary particles. Secondly, when an observation is made, it generally involves a complex dynamical interaction among many worldlines as a function of the evolution parameter $\tau$. However, the equations governing spacetime relaxation, geodesic relaxation, crystallization, and the additional dynamics of worldline interactions are all constructed to ensure that these processes ultimately decay, yielding a stable configuration of worldlines in the limit of $\tau\rightarrow \infty$. The outcome of an experiment can be inferred by the observer from this equilibrated, stable configuration of worldlines associated with the measurement apparatus. For instance, a typical pointer state (the position of a mechanical lever, a light being switched on or off, etc...) can be characterized by one or more coordinates of the form $(t,x,y,z)$ through conventional means—namely, spatial measurements with rulers and temporal readings from clocks. Crucially, although observers can record only the four spacetime coordinates $(t,x,y,z)$ associated with the equilibrated pointer state, they have no direct access to the $\tau$-dependent dynamics of worldlines, the relaxation of spacetime, or the value of $\tau$ itself. Hence, by collecting many outcomes—each described by coordinates $(t,x,y,z)$—observers can reconstruct a consistent picture of their world, one that agrees with the familiar classical four-dimensional spacetime. It is in this sense that we say a classical spacetime emerges from the perspective of an observer. Furthermore, as $\tau$ increases, the crystallization hypersurface $\Sigma(\tau)$ shifts further in the time dimension, leading to new regions of spacetime that reach equilibrium and to observations characterized by increasing values of the time coordinate $t$. In this sense, the observed increment of clock time $t$ indirectly probes the evolution parameter $\tau$, as is observed also for the coordinate time of $\Sigma(\tau)$ in Eq.~\eqref{crystallization}. This offers a natural explanation for the observation of the flow and arrow of time. 

Fig.~\ref{FIG0} schematically illustrates some of these principles. The magenta dots in Fig.~\ref{FIG0}a) represent four-dimensional events of measurement outcomes that have reached equilibrium. Most measurement outcomes are expected to quickly reach equilibrium, with dynamics restricted to the light blue region. But there can also be dynamics that stretch outside this region, far behind the crystallization hypersurface. For example, in the EPR experiment explored in Section~\ref{EPR} dynamics occur in a region of milliseconds behind the crystallization hypersurface. Importantly, from the perspective of an observer these events appear to belong to a traditional four-dimensional spacetime, which does not evolve as a function of $\tau$, as shown in Fig.~\ref{FIG0}b).

\subsection{Fundamental classicality}

Classicality is an ill-defined concept. It usually refers to determinism, realism, and locality in the context of standard spacetime, for example in standard classical mechanics or in general relativity. But, also higher-dimensional theories like five-dimensional general relativity are referred to as classical theories, such that the concept of classicality is not exclusively restricted to four-dimensional spacetime. In theories which abandon standard four-dimensional spacetime, the usual notions of determinism, realism and locality may not be applicable straight away. Nonetheless, similar concepts may be applied in such theories at a more fundamental level. Therefore, to avoid confusion, in this work we use the term ``fundamental classicality'' to refer to fundamental determinism, fundamental realism, and fundamental locality operating at a deeper level of reality beyond ordinary spacetime. More specifically, fundamental determinism means that each configuration of physical entities existing across four-dimensional spacetime at a value of $\tau+\Delta \tau$ depends in a deterministic way on the configuration at value $\tau$. Such fundamental determinism also implies fundamental reversibility. Fundamental realism broadly means that there are actually existing physical entities associated to particles, even before a measurement is performed as a function of $\tau$. And fundamental locality means that influences can only produce effects in a local region of four-dimensional spacetime as a function of $\tau$. One way to define this mathematically is to express that invariant intervals between any two physically existing entities must vary in a continuous way as a function of $\tau$, or: $\text{lim}_{\tau\rightarrow \tau_1} \Delta s^2(\tau)=\Delta s^2(\tau_1)$. In such a fundamentally classical theory, fundamental objects can be represented by well-defined and mutually exclusive pure states forming a simplex~\cite{Galley2023, Plavala2023}. Importantly, this does not rule out the emergence of ``non-classical'' features like superpositions, nonlocality and uncertainty from the perspective of a standard observer. In other words, a theory can be fundamentally classical, yet non-classical from the traditional perspective of standard four-dimensional spacetime.


\begin{figure}
\centering
\includegraphics[width=16.0cm]{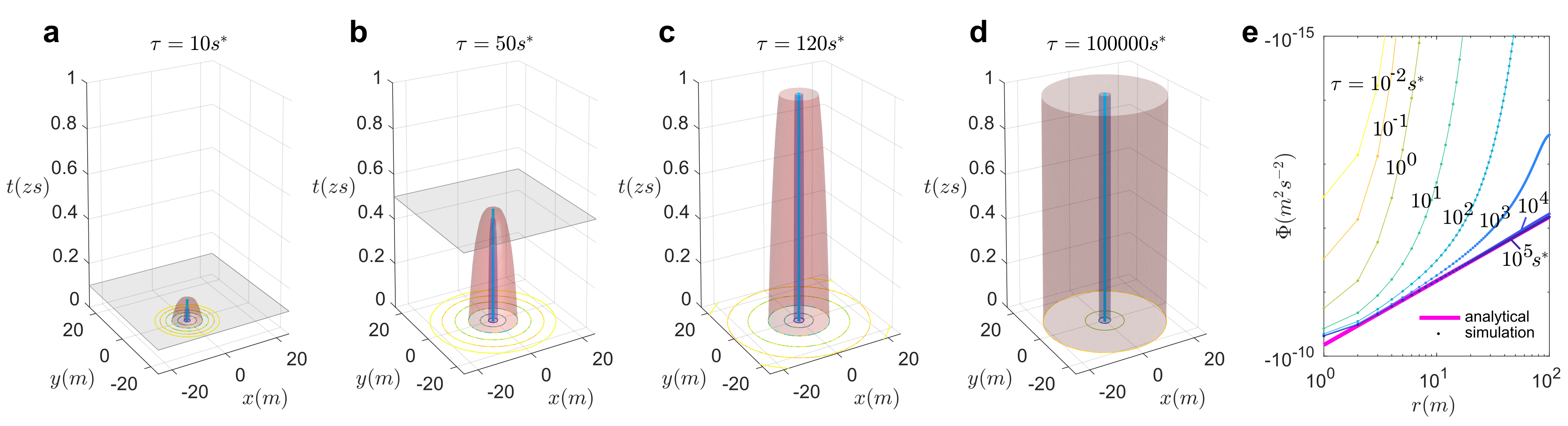}
\caption{Spacetime relaxation for a static mass in the weak-gravity approximation. \textbf{(a-d)} The gravitational potential $\Phi$ generated by a central mass (blue worldline) is displayed across a spacetime region spanning about 1 zs (i.e., $10^{-21}$ s) in the time dimension, with its evolution as a function of the parameter $\tau$ shown at $\tau= 10, 50, 120,$ and $10^5$ $s^*$. Key parameters include the mass $M=1$ kg, the spacetime relaxation rate $\kappa_{str}=1s^*/m^2$, and the crystallization rate $\beta=10^{-23}$ $s/s^*$. As the crystallization hypersurface $\Sigma(\tau)$ (grey surface) advances along the time axis, the gravitational influence of the mass extends into the time dimension. Around $\tau\approx 10^5$ $s^*$ the spacetime reaches its equilibrium, corresponding to a standard Newtonian spacetime. Potential contours are shown at  $-10^{-11.5}m^2s^{-2}$ (red) and $-10^{-10.5}m^2s^{-2}$ (purple). \textbf{(e)} The potential $\Phi(r,t=0)$ is shown as a function of the radial distance $r$ from the mass at different $\tau$ values, confirming that agreement with Newtonian gravity (magenta line) is reached around $\tau\approx 10^5$ $s^*$. See Methods~\ref{MethodsSimSTR} for more information.}\label{FIG1}
\end{figure}

\section{The weak gravity regime}

To analyze some basic principles of crystallizing spacetime, we adopt the regime of weak gravity. Here, we focus on recovering classical, Newtonian mechanics from the perspective of an observer. In the next sections, quantum phenomena will be addressed.

\subsection{Spacetime relaxation in the weak gravity regime}
 In the weak gravity regime Eq.~\eqref{MasterEq} reduces to (see Methods~\ref{MethodsSTRweak}):
\begin{equation}
  \kappa_{str}\frac{\partial \Phi(\tau)}{\partial \tau}= \nabla^2 \Phi(\tau) -4\pi G \rho(\tau)\label{weak},
\end{equation}
where $\Phi(\tau)$ is the $\tau$-dependent gravitational potential and $\rho(\tau)$ is the $\tau$-dependent mass density. In the limit for $\tau \rightarrow \infty$, where both $\Phi(\tau)$ and $\rho(\tau)$ reach an equilibrium in which $\partial \Phi(\tau) / \partial\tau=0$ and $\partial \rho(\tau) / \partial\tau=0$, we retrieve, as desired, the equation for Newtonian gravity:
\begin{equation}
   \nabla^2 \Phi =4\pi G \rho\label{NewtonianEquilibrium}.
\end{equation}
Fig.~\ref{FIG1} illustrates the concept of spacetime relaxation in the weak-gravity limit, with numerical simulations of Eq.~\eqref{weak} for a static mass $M=1$ kg, a spacetime relaxation rate $\kappa_{str}=1s^*/m^2$, and a crystallization rate $\beta=10^{23}s/s^*$ (see Methods~\ref{MethodsSimSTR} for details). For this static mass, geodesic relaxation is ignored and a straightforward particle worldline is considered in the past region with respect to the crystallization hypersurface $\Sigma(\tau)$ (blue line). Figs.~\ref{FIG1}a-c demonstrate how the gravitational effect of the mass gradually develops in the wake of the crystallization hypersurface $\Sigma(\tau)$ (grey surface). After about $\tau=10^5 s^*$, the spacetime reaches its equilibrium in agreement with Newtonian gravity (see Fig.~\ref{FIG1}d). In  Fig.~\ref{FIG1}e, the relaxation of the potential is compared to the analytical expectation for a point mass with Newtonian potential $\Phi(r)=-GM/r$ (magenta line). Here, parameters $\kappa_{str}$ and $\beta$ are chosen such that the dynamic region of spacetime in the past of the crystallization hypersurface spans only about 1 zs (i.e., $10^{-21}s$) in the time dimension, but these parameters can be tuned to make this dynamic region arbitrarily small.

\subsection{Geodesic relaxation in the weak gravity limit}\label{georelax}
\begin{figure}[h]
\centering
\includegraphics[width=16.0cm]{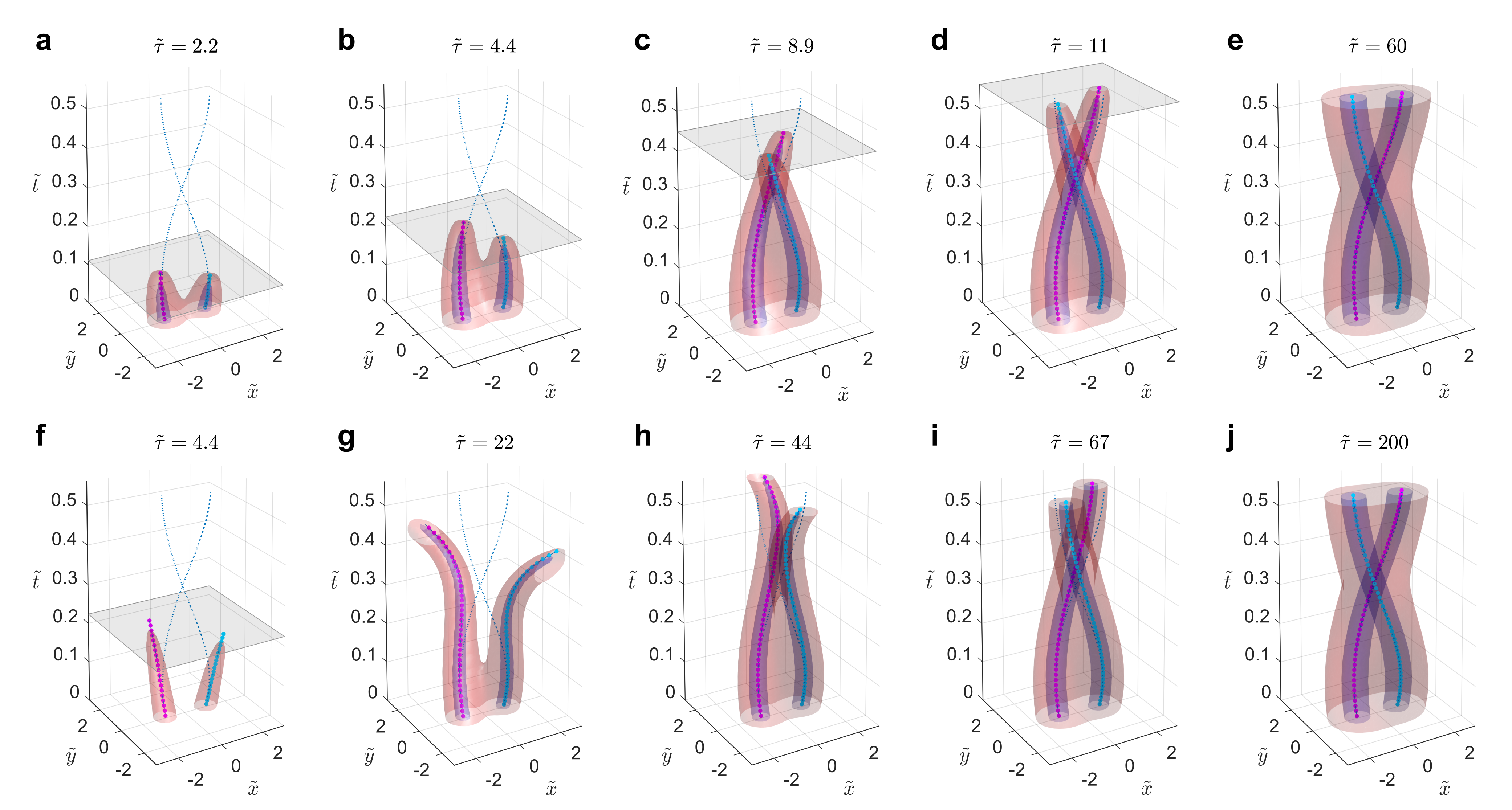}
\caption{Spacetime and geodesic relaxation in the weak-gravity regime. The Newtonian potential $\Phi$, originating from two equal masses $M$ (magenta and blue worldlines), is displayed at different values of the evolution parameter $\tau$, in normalized units. Key parameters include the mass $M=4\times10^9$ kg, the crystallization rate $\beta=0.01 s/s^*$, and the geodesic relaxation rate $\kappa_{geo}=1.25\times 10^{6} s^*$. As the crystallization hypersurface $\Sigma(\tau)$ (grey surface) advances in the forward time direction, the masses affect an increasing region of spacetime. Contours of the normalized potential are shown at values $-4\times 10^{-5}$ (purple) and $-2\times 10^{-5}$ (red). \textbf{(a-e)} In the case of fast spacetime relaxation, with $\kappa_{str}=1s^*/m^2$, worldlines quickly settle into equilibrium, closely following the expectation from standard Newtonian gravity (dotted lines). \textbf{(f-j)} In the case of slow spacetime relaxation, with $\kappa_{str}=10s^*/m^2$, worldlines first run straight in \textbf{(f)}. When the gravitational effect slowly builds up in \textbf{(g-i)}, worldlines are gradually drawn to each other. In both cases, at large values of $\tau$ (see \textbf{(e)} and \textbf{(j)}), the same equilibrium spacetime is obtained, which corresponds to a standard Newtonian spacetime. See Methods~\ref{sims2masses} for more information.}\label{FIG2} 
\end{figure}

For geodesic relaxation in the limit of weak gravity, the following approximation of Eq.~\eqref{geodesicrelaxation} is obtained for the time component $x^{0}$ and the spatial components $x^{\mu'}$ (see Methods~\ref{MethodsGRweak}):
\begin{align}\label{geoweak}
    &\kappa_{geo} \frac{\partial}{\partial\tau}(\frac{\partial^2 x^0}{\partial \lambda^2}) = -\Big[\frac{\partial^2 x^{0}}{\partial \lambda^2}\Big]  \nonumber\\
&\kappa_{geo} \frac{\partial}{\partial\tau}(\frac{\partial^2 x^{\mu'}}{\partial \lambda^2}) =-\Big[\frac{\partial^2 x^{\mu'}}{\partial \lambda^2} + \frac{\partial \Phi}{\partial x^{\mu'}}  \Big],
\end{align} 
where $\lambda$ is chosen to match with proper time. Consequently, in the limit of $\tau\rightarrow \infty$, Eq.~\eqref{geoweak} drives $\partial^2 x^{0}/\partial\lambda^2$ to zero, resulting as desired in a linearly increasing time coordinate as a function of $\lambda$. And, spatial acceleration $\textbf{a}=\frac{\partial^2}{\partial\lambda^2}(x^1,x^2,x^3)$ relaxes to $-\nabla \Phi$, in agreement with Newtonian gravity. Fig.~\ref{FIG2} illustrates the weak-gravity limit of geodesic relaxation in the case of two gravitationally interacting objects with equal masses $M=4\times10^9$ kg, separated initially by $3\times 10^4$ m and given initial velocities to produce a stable orbit in equilibrium (see Methods~\ref{sims2masses} for more information). Crystallization is implemented with $\beta=0.01 s/s^*$, while geodesic relaxation is realized by discretization of Eqs.~\eqref{geoweak} using $\kappa_{geo}=1.25\times 10^{6} s^*$. Fast geodesic relaxation is chosen such that spacetime relaxation and crystallization dominate the behaviour. Normalized coordinates $\tilde{x},\tilde{y},\tilde{z},\tilde{t}$, and $\tilde{\tau}$ are introduced as defined in Methods~\ref{sims2masses}. Regarding spacetime relaxation, two cases are explored. Figs.~\ref{FIG2}a-e illustrate fast spacetime relaxation by using a small value $\kappa_{str}=1 s^*/m^2$, for which spacetime quickly settles into its equilibrium configuration. Then, the dynamic region of spacetime corresponds to $\Delta \tilde{t} \approx0.5$ behind the crystallization hypersurface. In Figs.~\ref{FIG2}f-j, a ten times slower spacetime relaxation is considered, by using $\kappa_{str}=10 s^*/m^2$. Here, a larger region of $\Delta \tilde{t} \approx 10$ behind the crystallization hypersurface is affected. As expected, in both cases ultimately the same spacetime is obtained in the limit for $\tau\rightarrow \infty$, agreeing with Newtonian gravity (see Figs.~\ref{FIG2}e and j).

\subsection{Recovery of traditional Newtonian spacetime}

The above examples illustrate the basic principles of spacetime and geodesic relaxation in the weak-gravity limit. Now, we assess how this leads to the observation of a traditional Newtonian spacetime from the perspective of an observer. Firstly, it should be stressed that the spacetime and geodesic relaxation processes should be sufficiently fast compared to the advancement of the crystallization hypersurface to produce realistic scenarios. In other words, $\kappa_{str}$ and $\kappa_{geo}$ should be sufficiently small for a given value of $\beta$. For example, in the case of the macroscopic system with orbiting masses in Section~\ref{georelax} it is evident that the dynamics should be restricted to a region $\Delta \tilde{t} \ll1$ to be compatible with observations. This is required to explain that observers can accurately probe the trajectories of the orbiting masses at any value of $\tau$, within a time interval that is much shorter than the orbit time. As explained in Section~\ref{observation}, an observer can then record many events $(t,x,y,z)$ associated to the mass positions, finding that these are consistent with classical Newtonian mechanics. Furthermore, as $\tau$ increases, the observer collects more data characterized by an increasing value of the time coordinate, which is consistent with the experience of the flow of time. 

The same principle can be extended to a general crystallizing spacetime populated by $\tau$-dynamic worldlines, beyond the approximation of weak gravity. Then, following Eqs.~\eqref{MasterEq}, \eqref{crystallization}, and~\eqref{geodesicrelaxation}, a standard general relativistic spacetime emerges in equilibrium. However, due to the significant computational complexity involved, full simulations of these equations are beyond the scope of this work.

\section{Quantum nonlocality}\label{EPR}

\begin{figure}[h!]
\centering
\includegraphics[width=16.0cm]{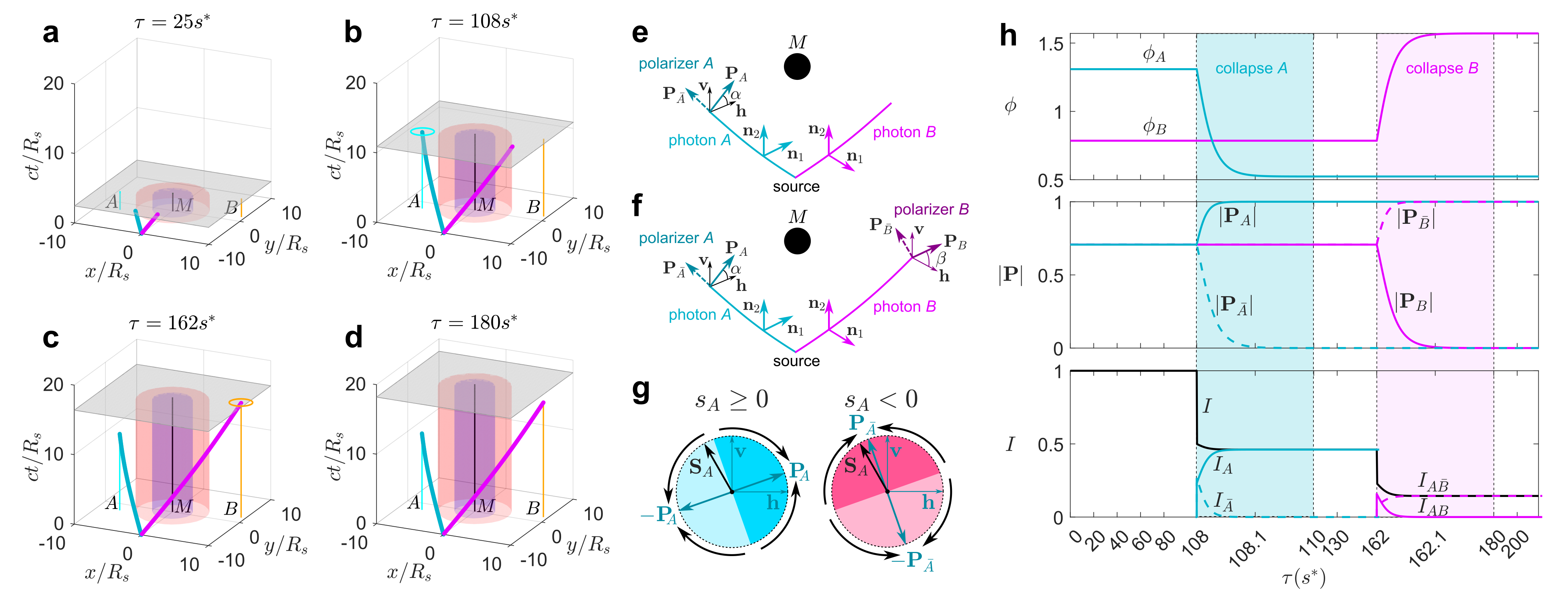}
\caption{A fundamentally classical model for EPR nonlocality within the crystallizing spacetime framework. \textbf{(a-d)} Photons $A$ and $B$ are sent to spacelike separated observers Alice and Bob. In the limit of fast spacetime relaxation, a fixed Schwarzschild spacetime is revealed behind the crystallization hypersurface $\Sigma(\tau)$ (grey surface). Evolution as a function of $\tau$ is illustrated for \textbf{(a)} the initial phase, \textbf{(b)} onset of interaction with Alice's polarizer, \textbf{(c)} onset of interaction with Bob's polarizer, and \textbf{(d)} the final phase. \textbf{(e-g)} Schematic illustration of interactions between photon and polarizer vectors after the onset of Alice's measurement \textbf{(e)} and Bob's measurement \textbf{(f)}, and of the collapse of Alice's polarizer vector \textbf{(g)}. \textbf{(h)} Evolution as a function of $\tau$ of the orientation $\phi$ of the polarizer state vector, the amplitudes $|\mathbf{P}|$ of the polarizer vectors, and the excitation intensities $I$ of loops along worldlines $A$ and $B$. Here, the case is shown of $n_1=0.65$, $n_2=0.76$, a pass for photon $A$, and a deflection for photon $B$. The cyan and magenta regions zoom in on the measurements of photons $A$ and $B$, respectively. The statistics and correlations of Alice's and Bob's outcomes agree with the expectations from quantum mechanics. See Supplementary Information and Methods~\ref{simsEPR} for more details.}\label{FIG3}
\end{figure}

In previous work, it has been demonstrated that quantum nonlocality occurring in an EPR experiment with a maximally entangled singlet state can be reproduced in a fundamentally classical way, in the absence of gravitational effects~\cite{Strubbe2023}. Here, within the crystallizing spacetime framework, we develop a model that reproduces more general EPR states and includes gravity. We focus on reproducing the outcomes of photons entangled through their polarization degrees of freedom, represented by the quantum state: 
\begin{equation}\label{singlet}
\left|\Psi \right\rangle=n_1 \left|H\right\rangle_{A} \left|H\right\rangle_{B} + n_2 \left|V\right\rangle_{A} \left|V\right\rangle_{B},
\end{equation}
where $\left|H\right\rangle$ and $\left|V\right\rangle$ represent horizontal and vertical polarization, $A$ and $B$ refer to Alice's and Bob's photons, and $n_1$ and $n_2$ are real numbers satisfying $n_1^2+n_2^2=1$. The main result is presented in Fig.~\ref{FIG3} for the case that $n_1=0.65$, $n_2=0.76$, photon $A$ passes while photon $B$ is deflected, with further details below, in Methods~\ref{simsEPR} and in Supplementary Information.

Within the crystallizing spacetime framework, the entangled photons are represented by two worldlines, $A$ and $B$, connected at the source in the event $e_S$. Worldlines $A$ and $B$ are emitted under an angle of 90° and travel respectively towards Alice and Bob, positioned left and right of a mass $M$  (see Figs.~\ref{FIG3}a-d). For simplicity, the regime of fast spacetime relaxation is adopted, by using a sufficiently small $\kappa_{str}$, such that a fixed gravitational background (with Schwarzschild metric) and fixed geodesics are revealed behind the hypersurface $\Sigma(\tau)$ as a function of $\tau$. The photon worldlines are characterized at each point by two orthogonal polarization vectors, $\mathbf{n}_1=n_1\mathbf{h}$ and $\mathbf{n}_2=n_2\mathbf{v}$, where $\mathbf{h}$ and $\mathbf{v}$ represent horizontal and vertical vectors orthogonal to the propagation direction $\mathbf{k}$ (see Fig.~\ref{FIG3}e). At the emission event $e_S$, we choose $\mathbf{v}_A=\mathbf{v}_B$ orthogonal to the equatorial plane, while $\mathbf{h}_A$ and $\mathbf{h}_B$ lie in the equatorial plane at an angle of $\pi/2$ with respect to each other. As the worldlines crystallize, all vectors are parallel-transported along the corresponding worldline.

Measurements are performed by Alice and Bob in spacelike separated regions using polarizing beam splitters $P_A$ and $P_B$ oriented at angles $\alpha$ and $\beta$ with respect to the $\mathbf{h}$-axis, respectively. Without loosing generality, we assume that the interaction event $e_A$ with Alice's polarizing beam splitter occurs first as a function of $\tau$, at $\tau=\tau_A$ (see Fig.~\ref{FIG3}b). When the worldline of photon $A$ interacts with Alice's polarizing beam splitter, the polarization vectors $\mathbf{n}_1$ and $\mathbf{n}_2$ couple to the polarizer vectors $\mathbf{P}_{A}(\tau)$ (oriented parallel to $\mathbf{1}_A$ at the angle $\alpha$) and $\mathbf{P}_{\bar{A}}(\tau)$ (oriented along $\mathbf{1}_{\bar{A}}$ at the angle $\alpha+\pi/2$) (see Fig~\ref{FIG3}e). To determine the outcome of this interaction, we consider closed interaction loops running along both photon worldlines $A$ and $B$, e.g. from the source to polarizer $P_A$, back to the source, then to polarizer $P_B$, and finally returning back to the source (see Supplementary Information for more details). Each loop is characterized by an excitation intensity $I$, corresponding to the amplitude of a wave with unit excitation amplitude after traveling across the entire loop as a function of $\tau$, according to the wave equation:
\begin{equation}\label{wave}
    \frac{\partial^2 A(\lambda,\tau)}{\partial \tau^2}=v^2\frac{\partial^2 A(\lambda,\tau)}{\partial \lambda^2},
\end{equation}
where $\lambda$ is a suitably chosen affine parameter along the loop and $v$ is the wave velocity. In analogy to a wave in standard spacetime, we can propose a solution of the form $A(\lambda,\tau)=A_0\text{cos}(k^*\lambda-\omega^*\tau+\phi)$, satisfying $\omega^{*2}=v^2k^{*2}$. We choose the parameter $k^*$, which may depend on the position $\lambda$ along the worldline, such that this corresponds to a desired local wavelength $\lambda_{ph}$ and frequency $\nu_{ph}$, while the remaining parameter $\omega^*$ can be chosen freely. This means that the wave velocity $v$ can be chosen arbitrarily large. Therefore, we will further assume for simplicity that $I$ adjusts instantaneously as a function of $\tau$ to changing coupling efficiencies in distant regions of spacetime (see details in Supplementary Information). Furthermore, since the feedback amplitude does not depend on where you start on the loop, the intensity $I$ becomes a shared property along the worldline loop. As a result, the coupling with polarizer vectors $\mathbf{P}_{A}(\tau)$ (associated with a pass) and $\mathbf{P}_{\bar{A}}(\tau)$ (associated with deflection) gives rise to the following excitation intensities: 
\begin{align}
    &I_A(\tau)= |\mathbf{P}_A(\tau)|^2( {n}_1^2 \text{cos}^2(\alpha)+{n}_2^2 \text{sin}^2(\alpha)) \nonumber \\
  &I_{\bar{A}}(\tau)= |\mathbf{P}_{\bar{A}}(\tau)|^2({n}_1^2 \text{sin}^2(\alpha)+{n}_2^2 \text{cos}^2(\alpha)).
\end{align}
 At the onset of the interaction, the magnitudes of both polarizer vectors are identical, namely, $|\mathbf{P}_{A}(\tau_A)|=|\mathbf{P}_{\bar{A}}(\tau_A)|=1/\sqrt{2}$, meaning that the outcome is still undecided. However, as $\tau$ increases, the polarizer state, governed by the evolution of the polarizer state vector $\mathbf{S}_A(\tau)=\mathbf{P}_A(\tau)+\mathbf{P}_{\bar{A}}(\tau)$, collapses either towards a pass state ($\pm\alpha$) or to a deflect state ($\pm(\alpha+\pi/2)$):
\begin{equation}\label{EPRevolutionlaw2}
    \frac{\partial \textbf{S}_{A}(\tau)}{\partial \tau} = \kappa_{meas} s_{A}(\tau)(\textbf{S}_{A}(\tau) \cdot \textbf{P}_{A})(\textbf{P}_{A} - (\textbf{S}_{A}(\tau) \cdot \textbf{P}_{A})\textbf{S}_{A}(\tau) ),
\end{equation}
with $\kappa_{meas}$ determining the relaxation rate of the collapse process, and with $s_A(\tau)=\text{sgn}(I_A(\tau)-r_A|\mathbf{P}_A(\tau)|^2)$ deciding the outcome based on a random hidden variable $r_A$ between 0 and 1 (as illustrated in Fig.~\ref{FIG3}g). The resulting probabilities for a pass or a deflection are:
\begin{align}
   & p(A)={n}_1^2 \text{cos}^2(\alpha)+{n}_2^2 \text{sin}^2(\alpha)\nonumber\\
    &p(\bar{A})={n}_1^2 \text{sin}^2(\alpha)+{n}_2^2 \text{cos}^2(\alpha),
\end{align}
as expected from quantum theory. When the collapse of photon $A$ is completed, $|\mathbf{P}_A|=1$ and $|\mathbf{P}_{\bar{A}}|=0$ in the case of a pass (as is illustrated in the cyan region of Fig.~\ref{FIG3}h), while $|\mathbf{P}_A|=0$ and $|\mathbf{P}_{\bar{A}}|=1$ in the case of deflection.

Finally, photon $B$ continues its path towards Bob's measurement station and interacts at $\tau=\tau_B$ with polarizer $P_B$ (see Figs.~\ref{FIG3}c and f). In the case that photon $A$ passed, only vector $\mathbf{P}_A=\mathbf{1}_A$ can be excited at side $A$, such that the following intensity contributions arise due to coupling with polarizer vectors $\mathbf{P}_B$ and $\mathbf{P}_{\bar{B}}$:
\begin{align}\label{BcurrentsApass}
   & I_{AB}(\tau)=|\mathbf{P}_B(\tau)|^2({n}_1\text{cos}(\alpha)\text{cos}(\beta) +{n}_2\text{sin}(\alpha)\text{sin}(\beta))^2 \nonumber\\
 &I_{A\bar{B}}(\tau)=|\mathbf{P}_{\bar{B}}(\tau)|^2 (-{n}_1\text{cos}(\alpha)\text{sin}(\beta) +{n}_2\text{sin}(\alpha)\text{cos}(\beta))^2.
 \end{align}
Similarly, also the intensities $I_{\bar{A}B}(\tau)$ and $I_{\bar{A}\bar{B}}(\tau)$ can be calculated. The outcome of the interaction between photon $B$ and polarizer $P_B$ is determined by an evolution law for the polarizer state vector $\mathbf{S}_B(\tau)=\mathbf{P}_B(\tau)+\mathbf{P}_{\bar{B}}(\tau)$ similar as in Eq.~\eqref{EPRevolutionlaw2}, but where $s_B(\tau)=\text{sgn}(I_{AB}(\tau)-r_B|\mathbf{P}_B(\tau)|^2)$ in case that $A$ passed and $s_B(\tau)=\text{sgn}(I_{\bar{A}B}(\tau)-r_B|\mathbf{P}_B(\tau)|^2)$ in the opposite case, with a random number $r_B$ between 0 and 1 (see magenta region in Fig.~\ref{FIG3}h). Since the sign of $s_B(\tau_B)$ determines the outcome for photon $B$ we obtain the following probabilities:
\begin{align}\label{Bprobabilities}
    p(A,B)&=({n}_1\text{cos}(\alpha)\text{cos}(\beta) +{n}_2\text{sin}(\alpha)\text{sin}(\beta))^2 \nonumber\\
 p(A,\bar{B})&=(-{n}_1\text{cos}(\alpha)\text{sin}(\beta) +{n}_2\text{sin}(\alpha)\text{cos}(\beta))^2\nonumber \\
 p(\bar{A},B)&=(-{n}_1\text{sin}(\alpha)\text{cos}(\beta) +{n}_2\text{cos}(\alpha)\text{sin}(\beta))^2 \nonumber \\
p(\bar{A},\bar{B})&=({n}_1\text{sin}(\alpha)\text{sin}(\beta) +{n}_2\text{cos}(\alpha)\text{cos}(\beta))^2,
\end{align}
in agreement with the predictions of quantum mechanics. 

This demonstrates that the nonlocal correlations of the considered EPR experiment are accurately reproduced in a fundamentally classical fashion. Namely, the presented EPR model is seamlessly embedded within a crystallizing four-dimensional spacetime, requiring no complex numbers or abstract spaces, with influences traveling deterministically along worldlines as a function of $\tau$, and with polarization degrees of freedom being influenced solely by their immediate surroundings. Therefore, the need for mysterious action-at-a-distance, which is assumed in standard quantum mechanics, is eliminated, and an implementation of Costa de Beauregard's zigzag action is achieved. As explained earlier, standard observers Alice and Bob can record the outcomes of their measurements in the form of four-dimensional coordinates $(t,x,y,z)$ associated with classical pointer states, which agree with a traditional view of spacetime. The key insight of this result is that, although the nonlocal correlations they observe from the perspective of standard spacetime defy classical intuition—seemingly requiring "spooky action at a distance"— the proposed model provides an underlying mechanism to make sense of these observations.

\section{Double-slit interference of a massive particle}\label{NeutronDS}

\begin{figure*}[ht!]
\centering
\includegraphics[width=16.0cm]{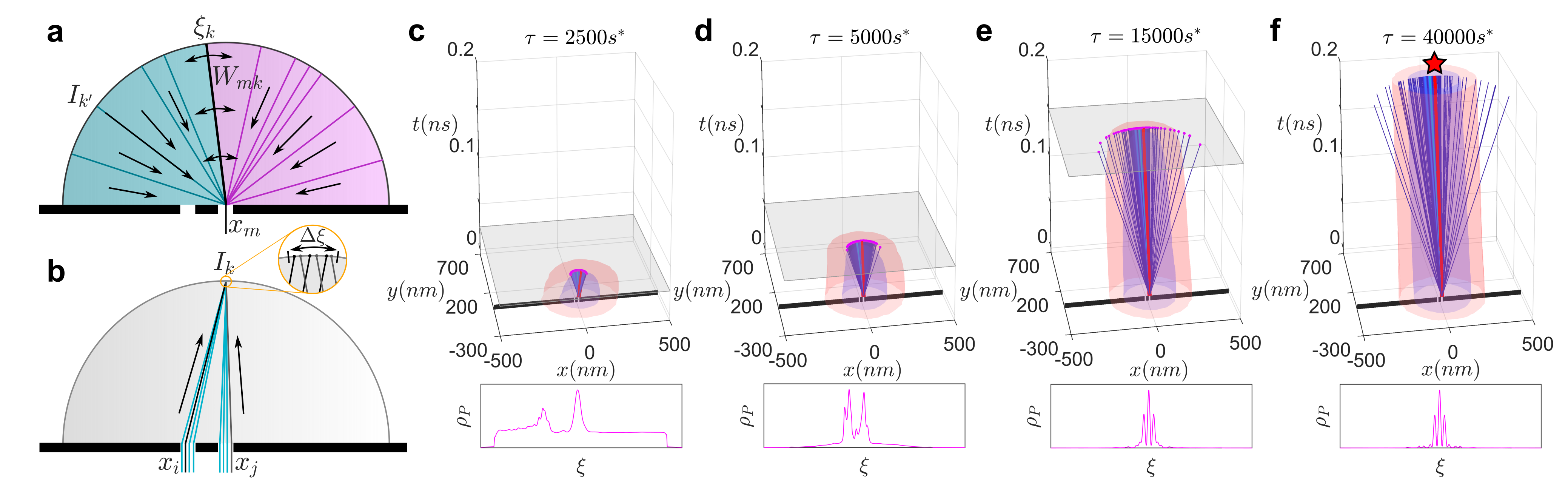}
\caption{A fundamentally classical model for double-slit interference of a massive particle within the crystallizing spacetime framework. \textbf{(a-b)} Schematic illustration of influences traveling across worldlines as a function of $\tau$, altering the spatio-temporal organization of worldlines. \textbf{(c-f)} Simulation of a neutron with $\lambda_{dB}=2$ nm passing a double slit aperture with width $W=10$ nm and separation $D=25$ nm. The worldline configuration and the corresponding density $\rho_P$ (bottom) is shown at different values of the evolution parameter $\tau$. The gravitational potential adjusts to the changing direction of the momentum-carrying worldline (highlighted in red). See Methods~\ref{simsDS} for more information.}\label{FIG4}
\end{figure*}

In prior work, it was shown that single-photon double-slit interference can be reproduced in a fundamentally classical way, in a scenario excluding gravitational effects~\cite{Strubbe2022}. In this study, we extend this concept by developing a fundamentally classical model for double-slit interference of a massive particle within the crystallizing spacetime framework, including its gravitational effect. The key findings are presented in Fig.~\ref{FIG4}, with additional details provided below and in Methods~\ref{simsDS}.

We consider a neutron with a de Broglie wavelength $\lambda_{dB}=2$ nm and velocity $v_P=2,000$ m/s in the non-relativistic regime, incident on a double-slit aperture. The slit width $W=10$ nm and separation $D=25$ nm are intentionally chosen to be smaller than in typical experiments~\cite{Zeilinger1988} to increase the diffraction angle for improved clarity. Each neutron is modeled by $N$ worldlines that approach the double slit aperture. Beyond the slits, each incident worldline at position $x_m$ branches into $M$ new worldlines $W_{mk}$, with $k=1,..,M$. As the hypersurface $\Sigma(\tau)$ progresses forward in time, these worldlines organize in a specific manner as a function of $\tau$. Only one randomly selected worldline (referred to next as the momentum-carrying worldline), comprising one incident worldline at the double-slit aperture and one connected outgoing worldline, carries energy and momentum, and ultimately determines the experimental outcome. 

Let us examine the situation at a specific value of $\tau$ in the reference frame of the double slit. We focus on the bundle of $M$ worldlines originating from the position $x_{m}$, to which the momentum-carrying worldline belongs (see Fig.~\ref{FIG4}a). The endpoints of these worldlines are distributed along a semi-circle of radius $L=v_Pt=v_P \beta \tau$, centered at $x_{m}$ and parametrized by the coordinate $\xi$. The density of worldlines along this semi-circle is denoted by $\rho_P(\xi)$. At a given point $\xi_k$, this density is calculated as $\rho_k=1/\Delta \xi_k$, where $\Delta \xi_k$ represents the spacing between worldline endpoints on the semi-circle. As illustrated in Fig.~\ref{FIG4}b, at each position $\xi_k$, an intensity $I_k=I(\xi_k)$ is determined by summing the contributions of all excitation loops $x_i \rightarrow \xi_{k'}\rightarrow \xi_{k''} \rightarrow x_j$, where $\xi_{k'}, \xi_{k''}$ lie within the elementary interval $[\xi_k,\xi_k+\Delta \xi]$:
\begin{equation}\label{Ik}
 I_k(\xi_k)=\sum_i \sum_j\frac{\sum_{k'=1}^{K_i} \sum_{k''=1}^{K_j} \text{cos}(\phi_{ik'}-\phi_{k''j})}{K_iK_j}
\end{equation}
where $K_i$ and $K_j$ are the number of worldlines originating respectively from $x_i$ and $x_j$. The accumulated phases are approximated by $\phi_{ik'}=2\pi l_{ik'}/ \lambda_{dB}$, where $l_{ik'}$ represents the spatial distance between positions $x_i$ and $\xi_{k'}$. Additionally, the coupling efficiencies $\xi_{k'}\rightarrow \xi_{k''}$ between neighboring worldline endpoints in this interval are assumed to be unity. Eq.~\eqref{Ik} indicates that the resulting excitation intensity at the position $\xi_k$ is obtained by summing over all excitation loops within the interval $\Delta \xi$, normalized by the number of participating loops. All information in Eq.~\eqref{Ik} is available in a fundamentally local way at position $\xi_k$ through waves propagating along worldlines, similar as described by Eq.~\eqref{wave} and as illustrated in Figs.~\ref{FIG4}a-b. These waves are excited in-phase at the double-slit aperture as a function of $\tau$, to reproduce the uniform initial phase of the incoming particle in this case. By making the approximation that $\text{cos}(\phi_{ik'}-\phi_{k''j})\approx \text{cos}(\phi_{ik}-\phi_{kj})$ holds for all $k'$ and $k''$ in this interval, Eq.~\eqref{Ik} simplifies to: 
\begin{equation}\label{Ikapprox}
 I_k(\xi_k)\approx \sum_i \sum_j\text{cos}(\phi_{ik}-\phi_{kj}),
\end{equation}
which is independent of the worldline density $\rho_P(\xi_k)$. Eq.~\eqref{Ikapprox} matches the standard path integral expression $|\sum_i e^{i \phi_{ik}}|^2$ of quantum mechanics, and is therefore proportional to the desired density $\rho_P(\xi_k)$ of worldlines. To reproduce this density, the spatial angle $\alpha_k$, which determines the velocity components $v_{k,x}=v_P \text{sin}(\alpha_k)$ and $v_{k,y}=v_P \text{cos}(\alpha_k)$ of worldline $W_{mk}$ at position $x_m$, is adjusted as a function of $\tau$ based on the excitation intensities $I_{k'}$:
\begin{align} \label{deltaalpha}
     \kappa_{ds}\frac{\partial \alpha_k}{\partial \tau}=\frac{\sum_{k'=1}^{k}I_{k'} \Delta \xi_{k'}}{\sum_{k'=1}^{k}\rho_{k'} \Delta \xi_{k'}} -\frac{\sum_{k'=k}^{M}I_{k'} \Delta \xi_{k'}}{\sum_{k'=k}^{M}\rho_{k'} \Delta \xi_{k'}},
      \end{align}
with relaxation rate constant $\kappa_{ds}$. Eq.~\eqref{deltaalpha} indicates that the reorientation of each worldline is determined by the imbalance in the summed excitation intensity, normalized by the number of participating worldlines, coming from the left versus the right side of worldline $W_{mk}$ (see Fig.~\ref{FIG4}a). The excitation intensities $I_{k'}\Delta \xi_{k'}$ are transported as a function of $\tau$ along worldline $W_{mk'}$, from position $\xi_{k'}$ to the position $x_m$, through a fundamentally local process, either via a wave equation like Eq.~\eqref{wave} or through another suitable transport equation. As a result of Eq.~\eqref{deltaalpha}, the velocity components of worldline $W_{mk}$ evolve according to:
    \begin{align}
      \frac{\partial v_{k,x}}{\partial \tau}&=v_P \text{cos}(\alpha_k) \frac{\partial \alpha_k}{\partial \tau} \nonumber\\
       \frac{\partial v_{k,y}}{\partial \tau}&=-v_P \text{sin}(\alpha_k) \frac{\partial \alpha_k}{\partial \tau}.
\end{align}
Furthermore, each worldline $W_{mk}$ adjusts to the changing initial velocity at the base position $x_m$ as a function of $\tau$, following the geodesic relaxation dynamics described by Eq.~\eqref{geodesicrelaxation}. For simplicity, we assume that this relaxation occurs much faster than the progression of the crystallization hypersurface, such that in good approximation each worldline instantaneously adapts to the changing initial velocity as a function of $\tau$. As $\tau$ progresses, the worldlines reach a quasi-equilibrium in which $\frac{\partial \alpha_k}{\partial \tau}=0$ for all $k$. In this quasi-equilibrium state, Eq.~\eqref{deltaalpha} leads, for each $k$, to the condition :
\begin{align} \label{deltaalpha2}
     \frac{\sum_{k'=1}^{k}I_{k'} \Delta \xi_{k'}}{\sum_{k'=1}^{k}\rho_{k'} \Delta \xi_{k'}} = \frac{\sum_{k'=k}^{M}I_{k'} \Delta \xi_{k'}}{\sum_{k'=k}^{M}\rho_{k'} \Delta \xi_{k'}}.
      \end{align}
This condition can only be satisfied if $\rho_k$ is proportional to $I_k$, for all $k$. Therefore, in equilibrium, we obtain the desired relationship:
\begin{equation}
    \rho_P(\xi_k)=QI_k(\xi_k),
\end{equation}
where $Q$ is a constant automatically determined by the normalization condition $\sum_k\rho_k \Delta \xi_k\equiv M=Q \sum_k I_k \Delta \xi_k$. The corresponding probability distribution for finding a worldline at position $\xi_k$ is given by:
\begin{equation}\label{probDS}
   p(\xi_k)=\frac{\rho_P(\xi_k)}{M}=\frac{I_k(\xi_k)}{\sum_k I_k(\xi_k) \Delta \xi_k},
\end{equation}
which matches the expected probability distribution from standard quantum mechanics, knowing that $I_k$ approximates the path integral expression $|\sum_i e^{i \phi_{ik}}|^2$. For worldline bundles incident at a different position than $x_m$ within the double-slit aperture, a similar reasoning can be made. Since the probability for the worldline carrying energy-momentum to pass through the double-slit aperture at position $x_i$ is uniform, the overall probability for finding the energy-momentum worldline under a certain deflection angle $\alpha$ follows a similar probability distribution as Eq.~\eqref{probDS}. As the hypersurface $\Sigma(\tau)$ progresses in the time dimension, the phase differences in Eq.~\eqref{Ik} may vary, particularly in the near field. Consequently, the result in Eq.~\eqref{probDS} represents a quasi-equilibrium state that continuously evolves as a function of $\tau$. By choosing a sufficiently small value for $\kappa_{ds}$, we can ensure that the worldlines rapidly relax to equilibrium, allowing the distribution of worldlines at each value of $\tau$ to closely match the desired quantum probability distribution.

To illustrate this process for the case of a neutron passing through a double slit, Figs.~\ref{FIG4}c-f depict the evolution of a bundle of worldlines emanating from within the right slit at position $x_{m}=47.5 \mu$m as a function of $\tau$ (see Methods~\ref{simsDS} for simulation parameters). The grey surface represents the crystallization hypersurface $\Sigma(\tau)$. Worldlines (with $M=400$) are shown with colors indicating the density $\rho_P(\xi)$, which is also displayed at the bottom. The transition from the near field to the far field occurs around $\tau=10,000 s^*$. A momentum-carrying worldline (worldline 320 out of 400) is randomly selected and highlighted in red. The gravitational potential generated by this momentum-carrying worldline is governed by Eq.~\eqref{weak}. For clarity, only the gravitational effect of the neutron is shown, excluding that of the double-slit aperture. As the momentum-carrying worldline changes its spatio-temporal position behind the double slit as a function of $\tau$, particularly in the near field (see Figs.~\ref{FIG4}c-d), the associated gravitational potential dynamically adjusts and relaxes to these changes. Here, to emphasize the relaxation process, a relatively small relaxation rate constant $\kappa_{ds}$ has been chosen, such that in Fig.~\ref{FIG4}c the quasi-equilibrium is not yet reached. However, in the limit of small $\kappa_{ds}$ and of fast spacetime and geodesic relaxation, one can ensure that the worldline configuration accurately matches with the quantum mechanical prediction, and that the gravitational potential in the past of the crystallization hypersurface closely matches with the expected Newtonian potential of the momentum-carrying worldline, at each value $\tau$. 

A definite outcome is obtained when the momentum-carrying worldline interacts with the worldline of a detector at the event indicated by a star in Fig.~\ref{FIG4}f (note that the worldline representing the detector is not shown for clarity). Once this interaction occurs the momentum-carrying worldline becomes fixed, such that no further dynamics occur as a function of $\tau$. As explained earlier, an observer may record the outcome of the experiment in the form of four-dimensional coordinates $(t, x, y, z)$ associated with classical pointer states, agreeing with a traditional view of spacetime. Since the result of a single experiment is determined by a randomly selected momentum-carrying worldline, the interference pattern emerges only after many repetitions of the experiment, consistent with the predictions of quantum mechanics. The Born rule is thus reproduced in a fundamentally classical manner, without problems with measurement or collapse, and without requiring the use of complex numbers. The central insight of this result is that, although the interference pattern appears counterintuitive from the standpoint of standard spacetime—seemingly requiring concepts such as superposition and wavefunction collapse—the proposed model offers a clear underlying mechanism that accounts for the observation within a fundamentally classical framework.

Furthermore, since the momentum-carrying worldline relaxes to a geodesic and the associated gravitational potential dynamically adapts, the principle of conservation of energy and momentum is naturally upheld in equilibrium. This results in an intuitive framework where the transfer of recoil momentum between the neutron and the double slit occurs precisely at the expected location, namely, at the double slit itself. In equilibrium, the gravitational effect corresponds to a massive particle following a geodesic trajectory between the double slit to the detector. Similarly, the arrival time can be understood as the travel time along this geodesic path. In contrast, standard quantum theory and interpretations such as Bohmian mechanics face challenges in providing a clear explanation for energy-momentum conservation during measurement and in predicting the arrival time.

As a result, this theoretical model within the framework of crystallizing spacetime demonstrates that the interference pattern of a massive particle, along with its associated gravitational effect, can be fully explained in a fundamentally classical manner. Consequently, by accepting the concept of crystallizing spacetime, the phenomenon can be understood without requiring the quantization of gravity or invoking superpositions of spacetime.

\section{Discussion}

The crystallization spacetime framework presented in this work provides a compelling foundation for a unified theory of gravity and matter. On the one hand, the mechanisms of spacetime relaxation, crystallization, and geodesic relaxation, described by Eqs.~\eqref{MasterEq}, \eqref{crystallization} and ~\eqref{geodesicrelaxation}, show how a four-dimensional spacetime in agreement with standard general relativity (or Newtonian spacetime in the weak-gravity limit) can emerge from a dynamically evolving spacetime in the limit of $\tau\rightarrow\infty$. On the other hand, through additional evolution laws governing the dynamics and interactions of worldlines as a function of $\tau$, it has been demonstrated that also key quantum phenomena can be reproduced. Below, the key features of the framework are summarized and discussed, highlighting its potential as a candidate theory of quantum gravity.

Notably, the framework of crystallizing spacetime is fundamentally classical, and does not rely on complex numbers, abstract spaces, or quantum concepts like superposition or action-at-a-distance at its core. In the presented EPR and double-slit models, all interactions can be traced back to fundamentally deterministic, fundamentally local, and fundamentally realist interactions as a function of $\tau$. Gravitational effects are explained without the need for quantizing gravity or invoking superpositions of spacetime. This shows that, by moving beyond traditional spacetime and formulating a suitable alternative to quantum theory, one can challenge the mainstream assumption that gravity must be quantized. It offers a straightforward explanation for the so-called quantum-to-classical transition, considering that the framework is entirely fundamentally classical to begin with and since standard observers collect information in the form of classical events compatible with traditional spacetime. As a result, crystallizing spacetime advances an intuitive view of reality. It provides a fresh perspective on the Einstein-Bohr debates, suggesting that Einstein's quest for an intuitive completion of quantum mechanics may be achievable—albeit beyond the confines of traditional spacetime. Nevertheless, from an observer's perspective, experimental outcomes can still be interpreted within the standard quantum and spacetime frameworks, leading to the apparent emergence of quantum concepts like superposition and nonlocality. 

Additionally, in the crystallizing spacetime framework the assumption of irreducibly quantum matter is rejected, which interestingly enables unification with fundamentally classical gravity while preserving determinism at a fundamental level~\cite{Galley2023}. Crucially, the concept of matter being fundamentally classical does not conflict with Bell's theorem~\cite{Bell1964}, as the crystallizing spacetime framework departs from the conventional notion of spacetime implicit in Bell's argument. Instead, it permits influences to propagate freely across space and time as a function of $\tau$, fully relaxing the statistical independence assumption underlying Bell's theorem—where statistical independence assumes that measurement settings are uncorrelated with the hidden variables governing outcomes~\cite{Hance2022,Hossenfelder2020, Hooft2014,Cohen2020a,Wharton2020}. Crystallizing spacetime adopts a common future hypothesis to violate statistical independence and to explain nonlocal correlations, in contrast to a common past hypothesis which is usually associated with superdeterminism (see Supplementary Information)~\cite{Norsen2021}. More specifically, through dynamics occurring as a function of $\tau$, a form of retrocausality is implemented in a well-defined manner that is free from paradoxes. In contrast, retrocausality in a context of standard spacetime is typically poorly defined and does struggle with such paradoxes~\cite{Drezet2019,Hossenfelder2020,Kastner2017,Norsen2021,Wharton2020}.

Models have been developed in this work to replicate EPR nonlocality and double-slit interference in the crystallizing spacetime framework, two of the most essential quantum phenomena. These models rely entirely on $\tau$-dynamic worldlines, ensuring compatibility with the mechanisms of spacetime and geodesic relaxation. Furthermore, these models are based on the concept of excitation loops along worldlines. This introduces a novel perspective on what particles may actually be: bundles of dynamically interacting worldlines evolving as a function of $\tau$, with influences propagating both forward and backward in the time dimension, in the spirit of Costa de Beauregard's zigzag action~\cite{Costa}. As a result, these models serve as a foundation for an alternative interpretation of quantum mechanics that is compatible with the concept of crystallizing spacetime. The presented models for EPR nonlocality and double-slit interference effectively illustrate the core concepts of the framework, despite relying on various simplifying assumptions such as fast relaxation and weak gravity. In future work, the goal is to develop a more comprehensive theoretical formalism capable of fully replacing the standard quantum formalism without such simplifying assumptions.

The crystallizing spacetime framework provides a physical mechanism for quantum collapse through $\tau$-dynamics of worldlines.  Here, collapse takes place outside conventional spacetime, consistent with the analysis of Bancal \textit{et al.} of EPR experiments, which suggests that ``quantum correlations somehow arise from outside spacetime, in the sense that no story in space and time can describe how they occur''~\cite{Bancal2012}. As illustrated in the double-slit model, this collapse mechanism resolves issues related to energy-momentum conservation in quantum measurements, because it allows to readjust both the momentum-carrying worldline and its gravity as a function of $\tau$. As a result, when a particle is detected behind the double slit, its spacetime trajectory has in fact relaxed to a geodesic, with the required momentum transfer occurring at the double-slit aperture. Additionally, by relying on a single worldline that determines the outcome, this collapse mechanism avoids the traditional measurement problem. The Born rule of quantum mechanics then emerges naturally from fundamentally local interactions across spacetime as a function of $\tau$, over many runs of the same experiment. This also makes sense from the perspective of the arrival time problem of quantum mechanics. Namely, in standard quantum theory it is challenging to understand how a particle, e.g. emitted with an isotropic wave function, can be detected at a certain position and time that apparently matches with a straight line (or geodesic) from source to detector, as is observed experimentally, since this seems to suggest the particle "knew" all along in which direction to depart. Within the crystallizing spacetime framework, a natural explanation is provided for this observation. Here, there is a specific worldline that determines the measurement outcome and that, even though it can spatio-temporally readjust as a function of $\tau$, always relaxes to a geodesic. As a result, the arrival time of the measured particle indeed matches with a geodesic trajectory between source to detector. Furthermore, the possibility of a fundamentally deterministic explanation for quantum collapse could have significant implications for theories of consciousness. For instance, Orchestrated Objective Reduction (Orch OR) theory~\cite{Hameroff2014} relies on the assumption that quantum collapse is non-deterministic and non-computable, viewing it as a key mechanism for consciousness and free will. In contrast, this work proposes that quantum collapse may be fundamentally deterministic, implying that explanations of consciousness must extend beyond it.

The introduction of the evolution parameter $\tau$ in the crystallizing spacetime framework redefines the concept of causality. In this framework, fundamental causality is governed by $\tau$, while time $t$ merely serves as a coordinate of a physical dimension. This paradigm shift resolves the long-standing tension between the role of time as a dimension in general relativity and as an evolution parameter in quantum mechanics. Namely, measured events recorded by an observer are expressed as $(x,y,z,t)$, without direct reference to prior $\tau$-dynamics, agreeing with the classical notion of spacetime in general relativity. However, according to the crystallization mechanism, these events are registered with a sequentially increasing time coordinate, giving rise to the perceived flow of time. Therefore, from this perspective, due to the proportional relationship $t_{cryst}=\beta \tau$ as described by Eq.~\eqref{crystallization}, time $t$ can thus equally be interpreted as an evolution parameter, which is precisely what occurs in standard quantum mechanics. While extra-dimensional theories often rely on compactification or other techniques to conceal extra dimensions from detection, the crystallizing spacetime framework needs no such mechanism for concealment of the additional evolution parameter. Instead, it argues that $\tau$ is essential for solving the above-mentioned problems with time, and for explaining both the observation of the present moment and the flow of time.

Next, the crystallizing spacetime framework is situated in relation to other theories. Unlike the evolving block universe of Ellis \textit{et al.}~\cite{Ellis2010}, this framework explicitly incorporates an additional parameter $\tau$ to represent evolution, enabling $\tau$-dynamics of worldlines in the past region with respect to the crystallization front. This approach agrees, to some extent, with similar ideas by Carr~\cite{Carr2021}, who suggests an additional time to explain the observed flow of time, and by Cohen \textit{et al.}~\cite{Cohen2020a}, who propose that two notions of causality are needed to realize the step-by-step emergence of spacetime with retrocausal properties. Moreover, this framework shares features with earlier theories by Stueckelberg~\cite{Stueckelberg1941}, Horwitz and Piron~\cite{Horwitz}, and Land~\cite{ Land2020}, which also introduce an additional parameter $\tau$ alongside conventional four-dimensional spacetime. Crystallizing spacetime can also be viewed as a non-superdeterministic extension of 't Hooft's cellular automaton beyond standard spacetime. Here, superdeterminism refers to the violation of statistical independence due to a shared past cause, a concept often regarded as conspiratorial~\cite{Hooft2014,Norsen2021} (see Supplementary Information). Also Parisi-Wu theory is particularly noteworthy, as it offers an alternative formulation of quantum field theory using a parameter $\tau$ to drive stochastic evolution, enabling the evaluation of quantum field configurations in equilibrium~\cite{ParisiWu}. While the idea of combining crystallizing spacetime with Parisi-Wu theory to realize a quantum gravity theory based on a shared evolution parameter $\tau$ is appealing, Parisi-Wu theory depends on complex, higher-dimensional abstract spaces and is bound by the measurement problem. This renders it incompatible with the nature of crystallizing spacetime. Therefore, an alternative for quantum field theory, embedded directly in a four-dimensional crystallizing spacetime, yet possibly inspired by Parisi-Wu theory, may be a potential pathway towards a unified quantum gravity framework. There is also a conceptual similarity with the transactional interpretation of quantum mechanics, which also relies on information propagating forward and backward in time~\cite{Cramer1986}. However, the transactional interpretation does not clarify how such bidirectional signals make sense within the standard concept of spacetime. The advantage of crystallizing spacetime is that it explicitly uses an additional evolution parameter $\tau$ and relies on zigzag action along worldines to clarify precisely how influences propagate across spacetime.

Finally, the crystallizing spacetime framework makes several distinctive predictions for future tests of quantum gravity. First, it predicts negative outcomes for gravitationally induced entanglement in tabletop experiments~\cite{Marletto2017,Fuchs2024,Carney2019,Gasbarri2021}, diverging from common expectations in mainstream quantum gravity research. In the standard view of quantum gravity, a massive particle in a spatial superposition is thought to create a corresponding superposition of the gravitational field, which can then entangle a second particle also in a spatial superposition. However, within crystallizing spacetime, illustrated by the presented double-slit model, only a single definite worldline of a superposition actually gravitates. Furthermore, no interactions across spacetime are included in the spacetime relaxation mechanism of Eq.~\eqref{MasterEq} that could enable gravitationally induced entanglement. Consequently, no superposition of the gravitational field arises, ensuring that this particle cannot gravitationally induce entanglement on another particle in superposition. Notice that non-gravitational entanglement is achieved in the presented EPR model because this is explicitly implemented through interactions along worldlines as a function of $\tau$. Second, the framework predicts agreement with general relativity, in principle even at the smallest measurable scales, contrasting with the expectation from standard quantum gravity or stochastic gravity. However, to address potential issues related to singularities in black holes and at the Big Bang, the crystallizing spacetime framework may ultimately require further refinements at the smallest scales—not through quantization, but possibly through additional evolution laws or a discretization of the fundamental constituents of worldlines. And, third, in a double-slit experiment the framework suggests the possibility of determining which slit a massive particle passed through using classical gravitational measurements, without disrupting the interference pattern. This would imply a violation of the uncertainty principle of quantum mechanics. Importantly, such a result is not necessarily problematic, as there is no experimental basis to assume that the uncertainty principle applies to gravitational interactions, given the distinct characteristics and behaviors of gravity compared to other fundamental interactions~\cite{Boughn2009}. The above insights are relevant, as recent analyses of potential quantum gravity experiments predominantly consider theories in which gravity is either quantum mechanical or stochastic~\cite{Kryhin2025}, neglecting the possibility of a fundamentally classical framework.

\section{Conclusion}

This work explains how gravitational phenomena consistent with general relativity and essential quantum phenomena like nonlocality and double-slit interference can both emerge within a fundamentally classical framework of crystallizing spacetime. This framework thus provides a promising basis for a unified theory of matter and gravity, challenging the common view that gravity must be quantized. It envisions reality as a crystallizing four-dimensional spacetime governed by fundamentally classical laws, with causality driven by the external parameter $\tau$. These $\tau$-dynamics offer a physical mechanism for quantum collapse operating beyond standard spacetime, enabling to circumvent the measurement problem, and naturally explaining the observed flow of time. Experimentally, it predicts agreement with general relativity in quantum experiments, the absence of gravitationally induced entanglement, and the possibility to gravitationally extract which-way information without disrupting interference, diverging from mainstream expectations. Future investigations could aim to develop a more comprehensive formalism within this framework to replace the standard quantum formalism, ultimately identifying evolution laws governing the most elementary components of worldlines and their crystallization at the Planck scale. Such insights may deepen our understanding of how the observable universe emerges.

\section{Methods}\label{Methods}



\subsection{Spacetime relaxation in the weak-gravity limit}\label{MethodsSTRweak}

Starting from Eq.~\eqref{MasterEq}, we derive an approximation suited for the regime of weak gravity and non-relativistic velocities. In this limit, the metric $g_{\mu\nu}(\tau)$ can be approximated as a small perturbation on top of a $\tau$-independent flat spacetime:
 \begin{equation}\label{metricapprox}
     g_{\mu\nu}(\tau)=\eta_{\mu\nu}+h_{\mu\nu}(\tau),
 \end{equation}
 where $\eta_{\mu\nu}=$diag$(-1,1,1,1)$ and $|h_{\mu\nu}(\tau)|\ll 1$. Following standard linearization procedures, we define the trace-reversed perturbation $\bar{h}_{\mu\nu}=h_{\mu\nu}-\frac{1}{2}\eta_{\mu\nu}h$, with ${h\equiv h^\alpha}_{\alpha}$, for which the Einstein tensor simplifies to $G_{\mu\nu}=-\frac{1}{2}\Box \bar{h}_{\mu\nu}$, where $\Box$ is the d'Alembertian operator $\eta^{\sigma\rho}\partial_{\sigma}\partial_{\rho}$. Under these conditions, Eq.~\eqref{MasterEq} reduces to:
 \begin{equation}
 \kappa_{str}\frac{\partial h_{\mu \nu}(\tau)}{\partial \tau}= -\big(-\frac{1}{2}\Box \bar{h}_{\mu\nu}(\tau)-\frac{8\pi G}{c^4}T_{\mu\nu}(\tau)\big).
 \label{STRweakgravity1}
\end{equation}
Focusing on the 00-component (as other components are negligible in this regime) we relate Eq.~\eqref{STRweakgravity1} to the gravitational potential $\Phi(\tau)$ using $h_{00}(\tau)=\bar{h}_{00}(\tau)/2\equiv -2\Phi(\tau)/c^2$, and $T_{00}(\tau)=\rho(\tau) c^2$, yielding: 
 \begin{equation}
 \kappa_{str}\frac{\partial \Phi(\tau)}{\partial \tau}= \Big[\nabla^2-\frac{\partial^2}{c^2\partial t^2}\Big]\Phi(\tau)-{4\pi G}\rho(\tau).
 \label{WeakPhi}
\end{equation}
Restricting to cases where time derivatives of the potential can be ignored, Eq.~\eqref{WeakPhi} further simplifies to:
\begin{equation}
 \kappa_{str}\frac{\partial \Phi(\tau)}{\partial \tau}= \nabla^2\Phi(\tau)-{4\pi G}\rho(\tau).
 \label{WeakPhiNotimederiv}
\end{equation}
Given $\rho(\tau)$, Eq.~\eqref{WeakPhiNotimederiv} describes how the potential $\Phi(\tau)$ relaxes over spacetime as a function of $\tau$. In the equilibrium limit $\tau\rightarrow \infty$, the familiar equation for Newtonian gravity is recovered:
\begin{equation}
    \nabla^2 \Phi= 4 \pi G \rho.
\end{equation}

\subsection{Simulation of spacetime relaxation for a static source}\label{MethodsSimSTR}

In Fig.~\ref{FIG1}, we present numerical simulations of spacetime relaxation in the weak-gravity limit, for a static mass, based on Eq.~\eqref{WeakPhiNotimederiv} expressed in spherical coordinates:
\begin{equation}
\begin{split}
 \Phi(i,t,\tau+\Delta \tau) = \Phi(i,t,\tau) + \frac{\Delta\tau}{\kappa_{str}} \big( ((\Phi(i+1,t,\tau) -2 \Phi(i,t,\tau) + \Phi(i-1,t,\tau))/\Delta r^2 \\+ \frac{2} {r(i)} (\Phi(i+1,t,\tau) - \Phi(i-1,t,\tau)) /(2 \Delta r)) - 4\pi G \rho(i,t,\tau) \big),
 \end{split}
 \end{equation}
 where the radial coordinate $r(i)=(i-1)\Delta r$, and $t$ represents the considered time slice. At the origin, $r=0$ (corresponding to $i=1$) a boundary condition $d\Phi(r=0)/dr=0$ is applied due to symmetry, while at the outer boundary, $r=R$ (corresponding to $i=N$), an extrapolating boundary condition $\Phi(N,\tau)=\Phi(N-1,\tau)r(N-1)/r(N)$ is applied consistent with the expected $1/r$ decay of the potential at large distances. Key parameters in the simulation include $\Delta r=1$ m, $N=101$, $R=100$ m, $\Delta \tau=0.01$ s$^*$ and $\kappa_{str}=1 s^*/m^2$. The central mass $M=1$ kg is modeled as a Gaussian density distribution with $\sigma=1$ m. The simulation spans a total time of $T=1$ zs (i.e., $10^{-21}s$), with time step $\Delta t=T/100$. Both the potential and density are initialized to zero. Crystallization proceeds according to $t_{cryst}(\tau)=\beta \tau$, with $\beta=10^{23}s/s^*$. As $t_{cryst}$ advances to a new time slice $i$, the density of this slice is updated by:
\begin{equation}\label{Gaussian}
    \rho(i)=\frac{M}{(2\pi\sigma^2)^{(3/2)}}\text{exp}(-\frac{r(i)^2}{2\sigma^2}).
\end{equation}
The resulting gravitational potential $\Phi(r,t)$ is displayed in the $(x,y,t)$-space, at $z=0$, with isocontours marking the gravitational potential at values $-10^{-11.5}m^2s^{-2}$ (red) and $-10^{-10.5}m^2s^{-2}$ (purple).

\subsection{Geodesic relaxation in the weak-gravity limit}\label{MethodsGRweak}
Starting from the general equation for geodesic relaxation, Eq.~\eqref{geodesicrelaxation}, we focus on the regime of weak gravity and non-relativistic velocities. In this regime, in the summation over Christoffel symbols only terms with $\Gamma^{\mu}_{00}$ are non-negligible. Applying Eq.~\eqref{metricapprox}, we find that $\Gamma^{\mu}_{00}=-\frac{1}{2}\partial_{\mu}h_{00}$. Thus, Eq.~\eqref{geodesicrelaxation} simplifies to:
\begin{equation}
    \kappa_{geo}\frac{\partial}{\partial \tau}(\frac{\partial^2 x^{\mu}}{\partial \lambda^2}) = - \Big[\frac{\partial^2 x^{\mu}}{\partial \lambda^2} -\frac{1}{2}\partial_{\mu}h_{00}(\frac{\partial x^0}{\partial \lambda})^2 \Big].
\end{equation}
By choosing $\lambda$ as the proper time along the worldline, we approximate $\partial x^0 / \partial \lambda \approx c$. Furthermore, substituting $h_{00}(\tau)=-2\Phi(\tau)/c^2$ and ignoring time derivatives of the potential, this leads to simplified equations for the $x^0$ (or $ct$) component and spatial components $x^{\mu'}$:
\begin{align}\label{geolaws}
    &\kappa_{geo}\frac{\partial}{\partial \tau}(\frac{\partial^2 x^{0}}{\partial \lambda^2}) =-  \Big[\frac{\partial^2 x^{0}}{\partial \lambda^2}\Big] \nonumber \\
    &\kappa_{geo}\frac{\partial}{\partial \tau}(\frac{\partial^2 x^{\mu'}}{\partial \lambda^2}) =- \Big[\frac{\partial^2 x^{\mu'}}{\partial \lambda^2} + \frac{\partial \Phi}{\partial x^{\mu'}} \Big].
\end{align}

\subsection{Simulation of spacetime relaxation for two interacting masses}\label{sims2masses}
In Fig.~\ref{FIG2} of the main text, we present numerical simulations of spacetime relaxation in the weak-gravity limit for two gravitationally interacting masses, based on Eq.~\eqref{weak}. The following relaxation scheme is used:
\begin{equation}
\begin{split}
{\Phi}(i,j,k,t,\tau+\Delta \tau)= {\Phi}(i,j,k,t,\tau) + \frac{\Delta \tau}{\kappa_{str}} \big(({\Phi}(i+1,j,k,t,\tau) + {\Phi}(i-1,j,k,t,\tau)\\ + {\Phi}(i,j+1,k,t,\tau) + {\Phi}(i,j-1,k,t,\tau) + 
{\Phi}(i,j-1,k,t,\tau) +{\Phi}(i,j,k+1,t,\tau) +\\{\Phi}(i,j,k-1,t,\tau) 
- 6 {\Phi}(i,j,k,t,\tau))/\Delta x^2 -{4\pi G\rho}(i,j,k,t,\tau)\big).
\end{split}
\end{equation}
The grid spans a spacetime domain of size $40\times40\times40\times50$, with spatial resolution $\Delta x=2,500$ m, temporal resolution $\Delta t=5\times10^5$ s and evolution step $\Delta \tau=0.5\times 10^6 s^*$. Extrapolating boundary conditions are applied to approximate the expected $1/r$ decay of the potential at large distances. Crystallization is modeled by $t_{cryst}=\beta \tau$ using $\beta=0.01 s/s^*$. Since time runs proportional to proper time in the weak gravity limit, time is divided into 50 discrete slabs corresponding to fixed $x^0$ values. Geodesic relaxation is realized by discretization of Eqs.~\eqref{geolaws} using $\kappa_{geo}=1.25\times10^{6}s^*$. It was chosen to implement the change in acceleration $\Delta a_i/\Delta \tau= \Delta(x_{i+1}-2x_i+x_{i-1})/(\Delta \lambda^2 \Delta \tau)$ at time $t_i$ by updating the position $x_{i+1}$ of the forward worldline segment between $t_i$ and $t_{i+1}$ according to $\Delta x_{i+1}/\Delta \tau=\Delta\lambda^2(\Delta a_i/\Delta \tau)$. In Fig.~\ref{FIG2}(a) $\kappa_{str}=1s^*/m^2$ is used, while in Fig.~\ref{FIG2}(b) $\kappa_{str}=10s^*/m^2$ is applied. The simulation begins with two masses, each with $M=4\times 10^9$ kg, positioned symmetrically at $p_1=(-l_0,0,0)$ and $p_2=(+l_0,0,0)$, separated by $L=2 l_0$, representing 30\% of the spatial domain, and with $l_0=15,000$ m. Their initial velocities, $v_1=(0,+v_{init},0)$ and $v_2=(0,-v_{init},0)$, are set up with $v_{init}=\sqrt{\frac{GM}{2L}}$ to match the orbital speed of a stable binary system. The density of each mass is represented by a 3D Gaussian distribution as in Eq.~\eqref{Gaussian}, with standard deviation $\sigma=4,000$ m, centered on each mass position. The resulting motion is restricted to the $(x,y)$-plane. For comparison, the dotted line in Fig.~\ref{FIG2} shows the analytical expectation for two equal masses in stable Newtonian orbit. For clarity, $x$, $y$, $z$, $t$, $\tau$ and $\Phi$ are normalized into $\tilde{x}$, $\tilde{y}$, $\tilde{z}$, $\tilde{t}$, $\tilde{\tau}$ and $\tilde{\Phi}$ in Fig.~\ref{FIG2}. Spatial coordinates are normalized by $l_0$. Time is normalized by $t_0=\frac{2\pi L}{v_{init}^2}=\frac{48\pi}{\sqrt{G\rho_0}}$, the duration for each mass to complete a $2\pi$ orbit. The gravitational potential $\Phi$ is normalized by $\Phi_0=4\pi G\rho_0 l_0^2$, where $\rho_0=M/l_0^3$ is the normalization constant for the density. Finally, the evolution parameter $\tau$ is normalized to $\tilde{\tau}=\tau / (l_0^2 1s^*/m^2)$. These normalizations reduce Eq.~\eqref{weak} into:
\begin{equation}\label{normalized}
    \frac{\kappa_{str}}{1s^*/ m^2}\frac{\partial\tilde{\Phi}}{\partial \tilde{\tau}}=\tilde{\nabla}^2\tilde{\Phi}-\tilde{\rho}.
\end{equation}
To visualize the gravitational potential, we plot isocontours at values $\tilde{\Phi}=-0.004$ (purple) and $\tilde{\Phi}=-0.002$ (red).

\subsection{Simulation parameters of the EPR model}\label{simsEPR}

 In the simulation of Fig.~\ref{FIG3}, photon $A$ reaches Alice's measurement station positioned at $(x',y')=(-6.73 R_s,-2.45 R_s)$ after 1.08 ms, while photon $B$ reaches Bob's station at $(9.36 R_s, 1.88 R_s)$ after 1.62 ms. Since it takes about 2 ms to travel from Alice to Bob, in addition to the 1.08 ms to reach Alice, Alice and Bob perform measurements in spacelike separated regions. For reference, isocontours of the effective gravitational potential $\Phi_{eff}(r)=-\frac{GM}{r}(1+\frac{GM}{c^2r})$ are plotted at values 
 $\Phi_{eff}=-2\times 10^{16}$ $m^2/s^2$ (purple) and $\Phi_{eff}=-10^{16}$ $m^2/s^2$ (red). Eq.~\eqref{EPRevolutionlaw2} is solved to find the evolution of the polarizer state vectors $\mathbf{P}_A$ and $\mathbf{P}_B$. Coefficients $n_1=0.65$ and $n_2=0.76$ are chosen, corresponding to a non-maximally entangled quantum state. Orientations of the polarizing beam splitters are set to $\alpha=\pi/6$ and $\beta=0$. The measurement relaxation rate is set to $\kappa_{meas}=50$ $(s^*)^{-1}$. The hidden variable governing Alice's outcome is $r_A=0.2785$, corresponding to a pass for photon $A$, while $r_B=0.5469$, corresponding to a deflection for photon $B$. For clarity, the figure zooms in on those $\tau$-intervals where photon $A$ is measured (cyan region) and where photon $B$ is measured (magenta region). Notice that $\kappa_{meas}$ can be chosen to obtain arbitrarily fast collapse as a function of $\tau$.

\subsection{Simulation parameters of the double-slit model}\label{simsDS}

The simulation of Fig.~\ref{FIG4} uses a neutron mass $m_n=1.375\times 10^{-27}$kg, a de Broglie wavelength of 2 nm, and a velocity of $2,000$ m/s. The double slit is characterized by $W=10$ nm and $D=25$ nm. The crystallization rate constant is $\beta=10^{-14} s/s^*$. The simulated worldlines ($M=400$) behind the double slit originate from position $x_m=47.5$ $\mu m$. For the reorientation of worldlines a rate constant $\kappa_{ds}=10^4$ $(s^*)^{-1}$ is used, while for spacetime relaxation the used parameters are $\Delta \tau=1s^*$ and $\kappa_{str}=10^{38}s^*/m^2$. The gravitational potential is calculated with Eq.~\eqref{weak} on a $40\times40\times40\times 100$ grid of spacetime, spanning 1.4$\mu m$ in space and $2\times10^{-10}$s in time. Since the spatial resolution is only 35 nm, the mass density in each time slab is convoluted with a normal distribution with standard deviation 50 nm. The calculation of the excitation intensities $I_k$ relies on 42 worldlines starting from positions within the double-slit aperture. To visualize the gravitational potential, we plot isocontours at values ${\Phi}=-10^{-51}$ $m^2/s^2$ (purple) and ${\Phi}=-10^{-54}$ $m^2/s^2$ (red).

\section{Supplementary Information}
\subsection{Model for linearly polarized entangled photons}\label{MethodsEPR}

Here, we provide more details on reproducing, within the crystallizing spacetime framework, the quantum state considered in the main text:
\begin{equation}\label{singletgeneral}
\left|\Psi \right\rangle= n_1\left|H\right\rangle_{A} \left|H\right\rangle_{B} + n_2\left|V\right\rangle_{A} \left|V\right\rangle_{B},
\end{equation}
with real numbers $n_1$ and $n_2$. According to standard quantum theory, this state can be expanded in terms of the orthogonal measurement basis $(\left|\alpha\right\rangle, \left|\alpha^{\perp}\right\rangle)$ at the side of Alice, by using $\left|H\right\rangle_A=\text{cos}(\alpha)\left|\alpha\right\rangle-\text{sin}(\alpha)\left|\alpha^{\perp}\right\rangle$ and $\left|V\right\rangle_A=\text{sin}(\alpha)\left|\alpha\right\rangle+\text{cos}(\alpha)\left|\alpha^{\perp}\right\rangle$, and using a similar procedure at Bob's side:
\begin{eqnarray}\label{singletgeneralexpanded}
\left|\Psi \right\rangle= n_1 (\text{cos}(\alpha)\left|\alpha\right\rangle-\text{sin}(\alpha)\left|\alpha^{\perp}\right\rangle) (\text{cos}(\beta)\left|\beta\right\rangle-\text{sin}(\beta)\left|\beta^{\perp}\right\rangle) + \nonumber\\
n_2(\text{sin}(\alpha)\left|\alpha\right\rangle+\text{cos}(\alpha)\left|\alpha^{\perp}\right\rangle) (\text{sin}(\beta)\left|\beta\right\rangle+\text{cos}(\beta)\left|\beta^{\perp}\right\rangle),
\end{eqnarray}
which can be grouped as:
\begin{align}
\left|\Psi \right\rangle&= (n_1 \text{cos}(\alpha)\text{cos}(\beta) + n_2 \text{sin}(\alpha) \text{sin}(\beta))\left|\alpha\right\rangle\left|\beta\right\rangle\nonumber \\
&+(-n_1 \text{cos}(\alpha)\text{sin}(\beta) + n_2 \text{sin}(\alpha) \text{cos}(\beta))\left|\alpha\right\rangle\left|\beta^{\perp}\right\rangle \nonumber\\
&+(-n_1 \text{sin}(\alpha)\text{cos}(\beta) + n_2 \text{cos}(\alpha) \text{sin}(\beta))\left|\alpha^{\perp}\right\rangle\left|\beta\right\rangle \nonumber \\
&+(n_1 \text{sin}(\alpha)\text{sin}(\beta) + n_2 \text{cos}(\alpha) \text{cos}(\beta))\left|\alpha^{\perp}\right\rangle\left|\beta^{\perp}\right\rangle.
\end{align}
By applying Born's rule, the resulting outcome probabilities according to quantum mechanics are:
\begin{align}\label{Bprobabilities}
    p(A,B)=&| \langle\alpha\beta\left|\Psi \right\rangle|^2=({n}_1\text{cos}(\alpha)\text{cos}(\beta) +{n}_2\text{sin}(\alpha)\text{sin}(\beta))^2 \nonumber \\
 p(A,\bar{B})=&| \langle\alpha\beta^{\perp}\left|\Psi \right\rangle|^2=(-{n}_1\text{cos}(\alpha)\text{sin}(\beta) +{n}_2\text{sin}(\alpha)\text{cos}(\beta))^2 \nonumber \\
 p(\bar{A},B)=&| \langle\alpha^{\perp}\beta^{\perp} \left|\Psi \right\rangle|^2=(-{n}_1\text{sin}(\alpha)\text{cos}(\beta) +{n}_2\text{cos}(\alpha)\text{sin}(\beta))^2 \nonumber \\
p(\bar{A},\bar{B})=&| \langle\alpha^{\perp}\beta^{\perp}\left|\Psi \right\rangle|^2=({n}_1\text{sin}(\alpha)\text{sin}(\beta) +{n}_2\text{cos}(\alpha)\text{cos}(\beta))^2.
\end{align}
For example, in the case of $n_1=n_2=1/\sqrt{2}$, which corresponds to a maximally entangled quantum state~\cite{Aspect2002}, the outcomes simplify to:
\begin{align}\label{Pall}
&p(A,B)=p(\bar{A},\bar{B})=\frac{\text{cos}^2(\alpha-\beta)}{2} \nonumber\\
&p(A,\bar{B})=p(\bar{A},B)=\frac{\text{sin}^2(\alpha-\beta)}{2}.
\end{align}

\subsubsection{EPR model within crystallizing spacetime}
In the main text, a model within the framework of crystallizing spacetime is presented that reproduces the nonlocal correlations from Eq.~\eqref{Bprobabilities} in a fundamentally classical way. Essentially, this is achieved by relying on interactions occurring along worldlines as a function of the evolution parameter $\tau$. Hence, by using an explicit evolution parameter $\tau$, a consistent implementation of Costa de Beauregard's vision of zigzag action along worldlines~\cite{Costa} is obtained, whereas it remains unclear how other approaches like the transactional interpretation~\cite{Cramer1986} explain such retrocausal action in a standard spacetime without referring to an additional evolution parameter. Furthermore, the used strategy of action along worldlines as a function of $\tau$ makes the presented interpretation of quantum mechanics compatible with the concept of spacetime relaxation. In this way, it illustrates key principles for how gravitational and quantum mechanical phenomena can be unified in the crystallizing spacetime framework. Below, several aspects of the model are further elaborated.

\subsubsection{Photon worldlines}

In the EPR model and simulation, we adopt Schwarzschild coordinates $(t',r',\phi',\theta')$ for a system with a central mass $M=10M_{Sun}$ located at the origin. In the limit of fast spacetime relaxation, corresponding to small values of $\kappa_{str}$, we approximate that a standard Schwarzschild spacetime emerges in the past region of $\Sigma(\tau)$. Crystallization occurs according to $t_{cryst}=\beta \tau$ with a chosen crystallization rate of $\beta=10^{-5}s/s^*$. We also assume fast geodesic relaxation, corresponding to the limit of small values of $\kappa_{geo}$. Then, in good approximation, fixed null geodesics of photons are revealed as the hypersurface $\Sigma(\tau)$ shifts forward in the time dimension. 

We consider a source that produces two worldlines, $A$ and $B$, in the event $e_S=(t'_0,r'_0,\phi'_0,\theta'_0)=(0,10R_s,0,0)$ and at $\tau=\tau_S$, where $R_S$ is the Schwarzschild radius. These worldlines are constrained within the equatorial plane $\theta'=\pi/2$. Worldline $A$ is emitted at an angle $\phi'_{S}=-3\pi/4$ relative to the $r'$-axis, and worldline $B$ at $\phi'_{S}=+3\pi/4$. Each worldline is computed as a standard null geodesic, with initial conditions at the emission event $e_S$:
\begin{equation}
    \mathbf{k}=(\dot{t}'_0,\dot{r}'_0,\dot{\phi}'_0,\dot{\theta}'_0),    
\end{equation}
where:
\begin{eqnarray}
    \dot{t}'_0=\frac{\sqrt{(\dot{r}'_0)^2+(r'_0)^2(\dot{\phi}'_0)^2(1-R_s/r'_0)}}{c(1-R_s/r'_0)} \nonumber\\
    \dot{r}'_0= c\text{cos}(\phi'_{S})\sqrt{1-R_s/r'_0}\nonumber\\
    \dot{\phi}'_0=c\text{sin}(\phi'_{S})/r'_0\nonumber\\
    \dot{\theta}'_0=0,
\end{eqnarray}
with the dot representing derivatives  $\partial/\partial \lambda$ with respect to the affine parameter $\lambda$. The parameter $\lambda$ is incremented in steps of $\Delta \lambda=10^{-7}$, with a total of $N=20,000$ steps. At the emission event, vectors defining the orthogonal plane of the polarization vector are initialized as follows:
 \begin{eqnarray}
     \mathbf{h}=(0, \frac{r'_0 \dot{\phi}'_0 } {c\dot{t}'_0}, -\frac{\dot{r}'_0}{(1-R_s/r'_0)c \dot{t}'_0 r'_0},0) \nonumber \\
    \mathbf{v}=(0,0,0,1/r'_0).
 \end{eqnarray} 
In other words, vectors $\mathbf{v}_A$ and $\mathbf{v}_B$ are parallel and in the $\mathbf{1}_{\theta'}$ direction, while vectors $\mathbf{h}_A$ and $\mathbf{h}_B$ lie in the equatorial plane, at an angle of $\pi/2$ with respect to each other. Parallel transport of the vectors $\mathbf{h}$ and $\mathbf{v}$ is carried out under the condition $\nabla_{\mathbf{k}}\mathbf{h} =\nabla_{\mathbf{k}}\mathbf{v} =0$.

\subsubsection{Zigzag action along closed loops}

The EPR model developed within the crystallizing spacetime framework replicates nonlocal correlations by relying on fundamentally local interactions occurring as a function of $\tau$ in closed loops along both worldlines $A$ and $B$. Such loops may start, for example, at the source, lead towards the endpoint of photon $A$, back to the source, then towards the endpoint of photon $B$, and finally back to the starting point. But any point along the loop may serve as start point. Various loops may be considered that couple different photon worldlines with each other at the source and that couple photon worldlines to different polarizer vectors associated to either a pass or a deflect state.

To understand how polarizers and photons collapse into a state with the desired nonlocal correlations, we assume that each loop is characterized by an excitation intensity $I$ (a scalar). This excitation intensity $I$ corresponds to the feedback amplitude of a wave, that is excited with unit amplitude at a certain position on the loop, travels the entire loop, and returns to the same position as a function of $\tau$. This feedback process occurs in a finite interval $\Delta\tau$. Therefore, when an interaction occurs with a polarizer, or when the polarizer vectors representing the polarizer state change as a function of $\tau$, the intensity $I$ will be updated within a finite $\tau$-interval. Loops involving polarizer vectors associated with a particular outcome and that are characterized by a large excitation intensity will increase the probability of realizing this outcome. A detailed model for how an excited wave travels across a loop, carrying crucial information from distant regions of spacetime, is presented further below. 

Since parameters can be chosen such that this feedback occurs in an arbitrarily short interval $\Delta\tau$, in the following we will assume for simplicity that the excitation intensity $I$ at any point along the loop adapts instantaneously as a function of $\tau$ upon changing interactions and coupling coefficients at distant spacetime locations. 

\subsubsection{Measurement of photon $A$}
\begin{figure*}[ht!]
\centering
\includegraphics[width=10.0cm]{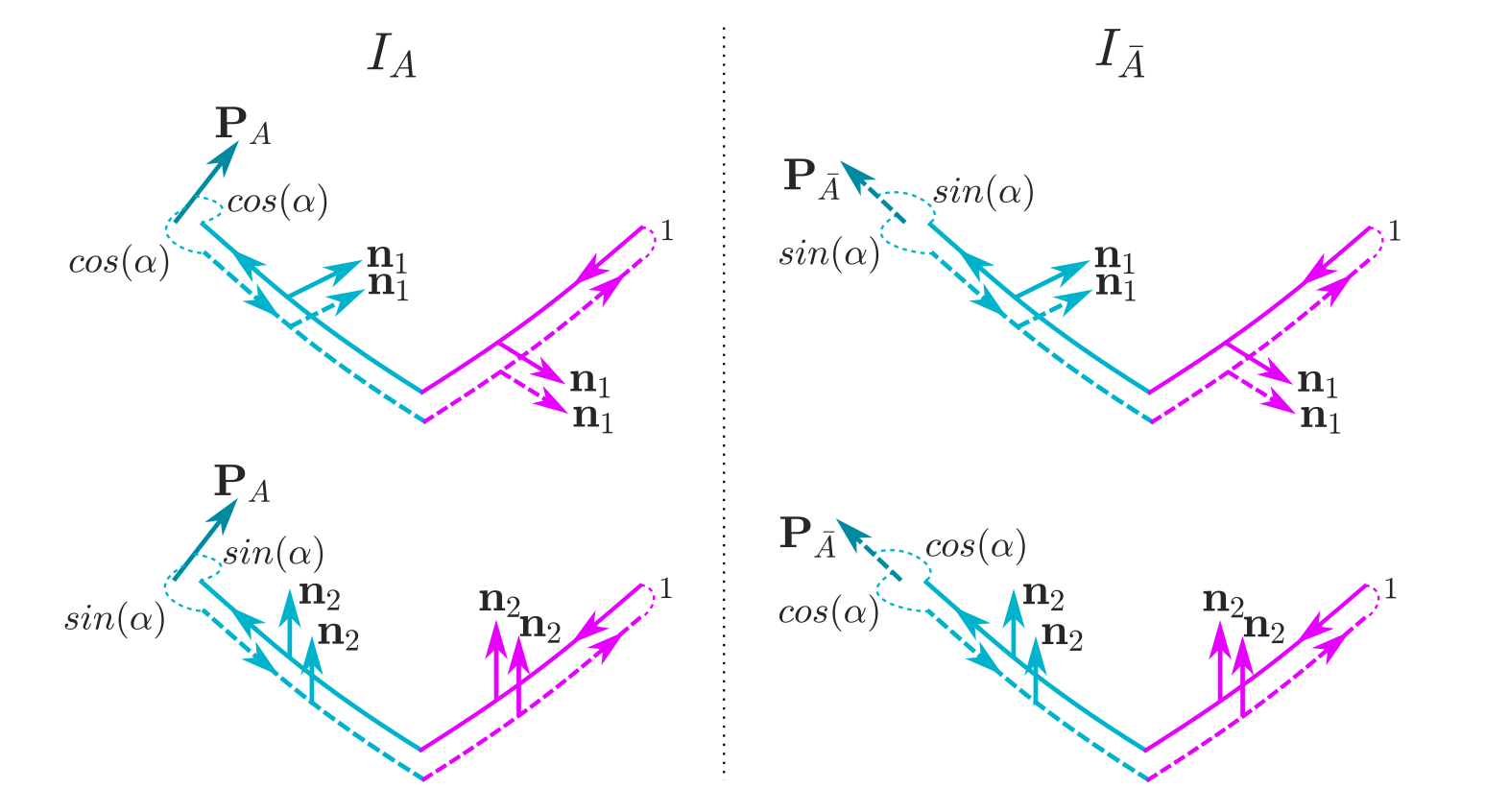}
\caption{Illustration of interaction loops with their coupling efficiencies contributing to the excitation efficiencies $I_A$ (left) and $I_{\bar{A}}$ (right), when photon $A$ interacts with Alice's polarizer, but before interaction with Bob's polarizer.}\label{FigSI1}
\end{figure*}

Firstly, we analyze the evolution of the system as a function of $\tau$ when photon $A$ interacts with Alice's polarizer at $\tau_A$, which we assume occurs before Bob's measurement at $\tau_B$ without loosing generality. For the considered quantum state (see Eq.~\eqref{singletgeneral}), $\mathbf{n}_1=n_1 \mathbf{h}$ and $\mathbf{n}_2=n_2 \mathbf{v}$. Alice's polarizer $P_A$ is characterized by polarizer vectors $\mathbf{P}_A(\tau)=|\mathbf{P}_A(\tau)|\mathbf{1}_A$ and $\mathbf{P}_{\bar{A}}(\tau)=|\mathbf{P}_{\bar{A}}(\tau)| \mathbf{1}_{\bar{A}}$. The unit vectors $\mathbf{1}_A$ and $\mathbf{1}_{\bar{A}}$, oriented respectively at angles $\alpha$ and $\alpha+\pi/2$ with respect to the $\mathbf{h}$-axis, represent states of the polarizer for which photon $A$ is respectively transmitted and deflected. Coupling between the polarization vectors (i.e., $\mathbf{n}_1$ and $\mathbf{n}_2$) and the polarizer vectors (i.e., $\mathbf{P}_A$ and $\mathbf{P}_{\bar{A}}$) produces an excitation intensity $I(\tau)$ in closed loops along both worldlines $A$ and $B$, starting and ending at polarizer $P_A$ (see Fig.~\ref{FigSI1}). In each loop, coupling efficiencies depend on the dot product of the involved polarization and/or polarizer vectors. For example, in the case of the vectors $\mathbf{n}_1$ and $\mathbf{P}_A$ the coupling efficiency is $\mathbf{n}_1 \cdot \mathbf{P}_A=|\mathbf{n}_1| |\mathbf{P}_A| \text{cos}(\alpha)$. And, at the end points of photon $B$ a self-coupling coefficient of 1 is taken for parallel vectors (e.g., $\mathbf{n}_1$ and $\mathbf{n}_1$), while the mutual coupling coefficient between orthogonal vectors is zero (e.g., $\mathbf{n}_1$ and $\mathbf{n}_2$). Here, at the source, only coupling between two vectors $\mathbf{n}_1$ (at sides $A$ and $B$) or between two vectors $\mathbf{n}_2$ is allowed.

Let us first examine the excitation of polarizer vector $\mathbf{P}_A(\tau)$, which is associated to a pass. Here, two loops can be identified (see Fig.~\ref{FigSI1}, left). A first loop, symbolized by $\mathbf{n}_1\xrightarrow{\text{cos}(\alpha)} \mathbf{P}_{A} \xrightarrow{\text{cos}(\alpha)} \mathbf{n}_1$, leads to a excitation intensity:
\begin{equation}
   I_{A,1}(\tau)= |\mathbf{P}_A(\tau)|^2 n_1^2\text{cos}^2(\alpha).
\end{equation}
This can be understood from the coupling between $\mathbf{n}_1$ and $\mathbf{P}_A(\tau)$, characterized by $\mathbf{n}_1 \cdot \mathbf{P}_A(\tau)=n_1 |\mathbf{P}_A(\tau)|\text{cos}(\alpha)$, followed by the inverse coupling between $\mathbf{P}_A(\tau)$ and $\mathbf{n}_1$, thus multiplying the coupling coefficient again by $\mathbf{n}_1 \cdot \mathbf{P}_A(\tau)=n_1 |\mathbf{P}_A(\tau)| \text{cos}(\alpha)$.
A second loop, $\mathbf{n}_2\xrightarrow{\text{sin}(\alpha)} \mathbf{P_{A}} \xrightarrow{\text{sin}(\alpha)} \mathbf{n}_2$, leads to:
\begin{equation}
   I_{A,2}(\tau)= |\mathbf{P}_A(\tau)|^2 n_2^2\text{sin}^2(\alpha).
\end{equation}
Loops where, for example, $\mathbf{n}_1$ excites $\mathbf{P}_A$, which in turn excites $\mathbf{n}_2$ do not contribute to the intensity considering the zero coupling efficiency between the orthogonal vectors $\mathbf{n}_1$ and $\mathbf{n}_2$. Therefore, the total excitation intensity associated with a pass state becomes:
\begin{equation}
    I_{A}(\tau)=I_{A,1}(\tau)+I_{A,2}(\tau)= |\mathbf{P}_A(\tau)|^2 (n_1^2\text{cos}^2(\alpha) + n_2^2\text{sin}^2(\alpha)).
\end{equation}
Secondly, we analyze the coupling with the polarizer vector $\mathbf{P}_{\bar{A}}(\tau)$. Similar as above, there are two loops (see Fig.~\ref{FigSI1}, right): $\mathbf{n}_1\xrightarrow{-\text{sin}(\alpha)} \mathbf{P_{\bar{A}}} \xrightarrow{-\text{sin}(\alpha)} \mathbf{n}_1$ and $\mathbf{n}_2\xrightarrow{\text{cos}(\alpha)} \mathbf{P_{\bar{A}}} \xrightarrow{\text{cos}(\alpha)} \mathbf{n}_2$, producing together an intensity:
 \begin{equation}
    I_{\bar{A}}(\tau)=I_{\bar{A},1}(\tau)+I_{\bar{A},2}(\tau)= |\mathbf{P}_{\bar{A}}(\tau)|^2 (n_1^2\text{sin}^2(\alpha) + n_2^2\text{cos}^2(\alpha)).
\end{equation}
The total excitation intensity is thus given by:
 \begin{equation}\label{totalcurrentA}
     I(\tau)=I_{A}(\tau)+I_{\bar{A}}(\tau)=|\mathbf{P}_A(\tau)|^2 (n_1^2\text{cos}^2(\alpha) + n_2^2\text{sin}^2(\alpha))+ |\mathbf{P}_{\bar{A}}(\tau)|^2 (n_1^2\text{sin}^2(\alpha) + n_2^2\text{cos}^2(\alpha)).
\end{equation}
The state of polarizer $P_A$, captured by the vector $\mathbf{S}_A(\tau)=\mathbf{P}_A(\tau)+\mathbf{P}_{\bar{A}}(\tau)$, evolves as a function of $\tau$ according to the following evolution law: 
\begin{equation}\label{EPRevolutionlaw2A}
    \frac{\partial \textbf{S}_{A}(\tau)}{\partial \tau} = \kappa_{meas} s_{A}(\tau) (\textbf{S}_{A}(\tau) \cdot \textbf{1}_{A})(\textbf{1}_{A} - (\textbf{S}_{A}(\tau) \cdot \textbf{1}_{A})\textbf{S}_{A}(\tau)).
\end{equation}
At the onset of the interaction, at $\tau=\tau_A$, the polarizer state is undecided, characterized by $\mathbf{P}_A(\tau_A)=\frac{1}{\sqrt{2}}\mathbf{1}_A$ and $\mathbf{P}_{\bar{A}}(\tau_A)=\frac{1}{\sqrt{2}}\mathbf{1}_{\bar{A}}$
such that $\mathbf{S}_A(\tau_A)=\frac{1}{\sqrt{2}}(\mathbf{1}_{{A}}+ \mathbf{1}_{\bar{A}})$. Eq.~\eqref{EPRevolutionlaw2A} can be reformulated in terms of the angle $\phi_A$ of the vector $\mathbf{S}_A(\tau)$ with respect to the $\mathbf{h}$-axis:
\begin{equation}\label{mmtconditionphi}
    \frac{\partial \phi_A(\tau)}{\partial \tau} = \frac{1}{2}\kappa_{meas} s_A(\tau) \text{sin}(2(\alpha-\phi_A(\tau))),
\end{equation}
by using $\mathbf{S}_A(\tau)=\text{cos}(\phi_{A}(\tau))\mathbf{h} + \text{sin}(\phi_{A}(\tau))\mathbf{v}$, $\mathbf{1}_A=\text{cos}(\alpha)\mathbf{h} + \text{sin}(\alpha)\mathbf{v}$, and noting that $\mathbf{n} \cdot \frac{\partial \mathbf{S}_A}{\partial \tau} = \frac{\partial \phi_A}{\partial \tau}$, with $\mathbf{n}=-\text{sin}(\phi_{A}(\tau))\mathbf{h} + \text{cos}(\phi_{A}(\tau))\mathbf{v}$. In essence, Eq.~\eqref{mmtconditionphi} forces the vector $\mathbf{S}_A$ to align either parallel with the polarizer vector $\mathbf{1}_A$, corresponding to a pass, or orthogonal to $\mathbf{1}_A$ (i.e., parallel to $\mathbf{1}_{\bar{A}}$), representing a deflection. More specifically, there are four possible scenarios to consider, depending on the sign of $s_A(\tau_{A})$ and on the angle between $\mathbf{S}_A$ and $\mathbf{1}_A$ at $\tau_A$, as illustrated in Fig. 3g of the main text. If $s_A(\tau_A)$ is positive, then $\mathbf{S}_A(\tau)$ is attracted towards $\textbf{1}_A$ if $\textbf{S}_A(\tau_A)\cdot \textbf{1}_A\geq 0$ (corresponding to $|\phi_{A}(\tau_A)-\alpha| \leq \pi/2$), and towards $-\mathbf{1}_A$ if $\textbf{S}_A(\tau_A)\cdot \textbf{1}_A < 0$ (corresponding to $|\phi_{A}(\tau_A)-\alpha| > \pi/2$). This outcome corresponds to photon $A$ passing through the beam splitter. However, if $s_A(\tau_A)$ is negative, $\mathbf{S}_A(\tau)$ is drawn towards $\mathbf{1_{\bar{A}}}$ if $\mathbf{S}(\tau_A)\cdot \mathbf{1_{\bar{A}}} \geq 0$ and towards $-\mathbf{1_{\bar{A}}}$ if $\mathbf{S}_A(\tau_A)\cdot \mathbf{1}_{\bar{A}}<0$, which corresponds to worldline $A$ being deflected. As the orientation of $\mathbf{S}_A(\tau)$ changes as a function of $\tau$, the magnitudes of the vectors $\mathbf{P}_A(\tau)$ and $\mathbf{P}_{\bar{A}}(\tau)$ change accordingly:
\begin{align}
   & \mathbf{P}_A(\tau)=(\mathbf{S}_A(\tau) \cdot \mathbf{1}_A) \mathbf{1}_A \nonumber\\
&\mathbf{P}_{\bar{A}}(\tau)=(\mathbf{S}_{A}(\tau) \cdot \mathbf{1}_{\bar{A}}) \mathbf{1}_{\bar{A}}.
\end{align}

The outcome of photon $A$ is either a pass or a deflection, determined by $s_A(\tau)=\text{sgn}(I_A(\tau)-r_A|\mathbf{P}_A(\tau)|^2)$ through the fundamentally local hidden variable $r_A$. This hidden variable is for simplicity taken to be a random number between 0 and 1. Given that $s_A(\tau_A)=\text{sgn}(I_A(\tau_A)-r_A|\mathbf{P}_A(\tau_A)|^2)$, we find that the probability for $s_A(\tau)\geq 0$ is:
\begin{equation}\label{outcomeApass}
    p(A)={n}_1^2 \text{cos}^2(\alpha)+{n}_2^2 \text{sin}^2(\alpha),
    \end{equation}
    while the probability for $s_A(\tau_A) < 0$ is:
    \begin{equation}\label{outcomeAdeflect}
    p(\bar{A})={n}_1^2 \text{sin}^2(\alpha)+{n}_2^2 \text{cos}^2(\alpha).
\end{equation}
During the collapse, for $\tau\geq \tau_A$, the sign of $s_A(\tau)$ remains the same, ensuring a consistent outcome. Since the sign of $s_A(\tau_A)$ determines the outcome, Eqs.~\eqref{outcomeApass} and \eqref{outcomeAdeflect} are also the probabilities for the outcomes of photon $A$, in agreement with quantum mechanics. 

Notice that, in the case of a pass for photon $A$, the polarizer vector $\mathbf{S}_A(\tau)$ collapses to $\mathbf{1}_A$, meaning that $|\mathbf{P}_A|$ becomes equal to $\mathbf{1}_A$ while $|\mathbf{P}_{\bar{A}}|$ becomes zero. The polarizer then collapses into a state that transmits a photon characterized by a polarization state aligned with the polarizer angle $\alpha$. The opposite state of the polarizer can then no longer be excited by the incident photon $A$. The excitation intensity expressed by Eq.~\eqref{totalcurrentA} then becomes:
\begin{equation}
     I(\tau)= n_1^2\text{cos}^2(\alpha) + n_2^2\text{sin}^2(\alpha).
\end{equation}
A similar reasoning can be made in the case that photon $A$ is deflected. Then, the intensity becomes:
\begin{equation}
     I(\tau)= n_1^2\text{sin}^2(\alpha) + n_2^2\text{cos}^2(\alpha).
\end{equation}

\subsubsection{Measurement of photon $B$}
Next, we analyze the evolution of the system when photon $B$ interacts with Bob's polarizing beam splitter in the interaction event $e_B$, when $\tau=\tau_B$. Similar as above, polarizer $P_B$ is characterized by two polarizer vectors $\mathbf{P}_B(\tau)=|\mathbf{P}_B(\tau)|\mathbf{1}_B$ associated with a pass and $\mathbf{P}_{\bar{B}}(\tau)=|\mathbf{P}_{\bar{B}}(\tau)|\mathbf{1}_B$ associated with deflection. The overall polarization state is characterized by $\mathbf{S}_B(\tau)=\mathbf{P}_B(\tau)+\mathbf{P}_{\bar{B}}(\tau)$. Now, all possible excitation pathways involving vectors $\mathbf{n}_1$, $\mathbf{n}_2$, $\mathbf{P}_A$, $\mathbf{P}_{\bar{A}}$, $\mathbf{P}_B$, and $\mathbf{P}_{\bar{B}}$ must be considered. However, assuming that photon $A$ is already fully collapsed when the interaction with polarizer $B$ is initiated, which is reasonable when considering a large value for $\kappa_{meas}$, a simplification can be made. Then, two different cases can be distinguished, depending on whether photon $A$ passes or is deflected.

\begin{figure*}[ht!]
\centering
\includegraphics[width=12.0cm]{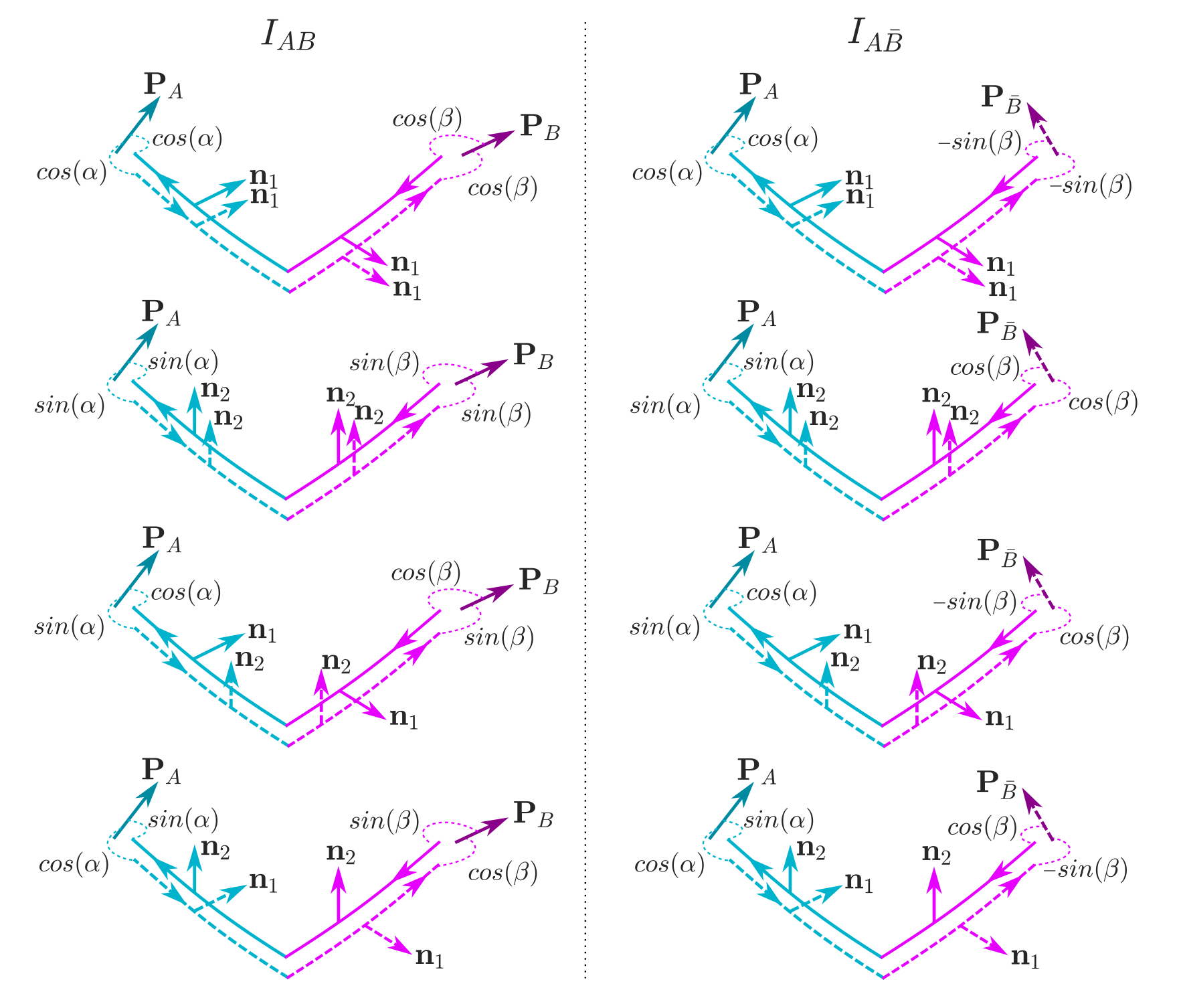}
\caption{Illustration of interaction loops with their coupling efficiencies contributing to the excitation intensities $I_{AB}$ (left) and $I_{A\bar{B}}$ (right), when photon $B$ interacts with Bob's polarizer, in the case of a pass for photon $A$.}\label{FigSI2}
\end{figure*}

In the case that photon $A$ passes, four excitation pathways couple to the polarizer vector $\mathbf{P}_B(\tau)$ and contribute to the excitation intensity $I_{AB}(\tau)$, summarized by (see Fig.~\ref{FigSI2}, left):
\begin{align}
&\mathbf{n}_1\xrightarrow{\text{cos}(\alpha)} \mathbf{P_{A}} \xrightarrow{\text{cos}(\alpha)} \mathbf{n}_1\xrightarrow{\text{cos}(\beta)} \mathbf{P_{B}} \xrightarrow{\text{cos}(\beta)} \mathbf{n}_1\nonumber \\
&\mathbf{n}_2\xrightarrow{\text{sin}(\alpha)} \mathbf{P_{A}} \xrightarrow{\text{sin}(\alpha)} \mathbf{n}_2\xrightarrow{\text{sin}(\beta)} \mathbf{P_{B}} \xrightarrow{\text{sin}(\beta)} \mathbf{n}_2\nonumber \\
&\mathbf{n}_1\xrightarrow{\text{cos}(\alpha)} \mathbf{P_{A}} \xrightarrow{\text{sin}(\alpha)} \mathbf{n}_2\xrightarrow{\text{sin}(\beta)} \mathbf{P_{B}} \xrightarrow{\text{cos}(\beta)} \mathbf{n}_1\nonumber \\
&\mathbf{n}_2\xrightarrow{\text{sin}(\alpha)} \mathbf{P_{A}} \xrightarrow{\text{cos}(\alpha)} \mathbf{n}_1\xrightarrow{\text{cos}(\beta)} \mathbf{P_{B}} \xrightarrow{\text{sin}(\beta)} \mathbf{n}_2.
\end{align}
The resulting intensity $I_{AB}(\tau)$ is:
\begin{equation}\label{AB}
    I_{AB}(\tau)=|\mathbf{P}_B(\tau)|^2({n}_1^2\text{cos}^2(\alpha)\text{cos}^2(\beta) +{n}_2^2\text{sin}^2(\alpha)\text{sin}^2(\beta)+2{n}_1{n}_2\text{cos}(\alpha)\text{cos}(\beta)\text{sin}(\alpha)\text{sin}(\beta)),
    \end{equation}
which can be simplified to:
\begin{equation}
      I_{AB}(\tau)=|\mathbf{P}_B(\tau)|^2({n}_1\text{cos}(\alpha)\text{cos}(\beta) +{n}_2\text{sin}(\alpha)\text{sin}(\beta))^2.
\end{equation}
Following a similar reasoning (see Fig.~\ref{FigSI2}, right), coupling with polarizer vector $\mathbf{P}_{\bar{B}}(\tau)$ gives the following contributions:
\begin{align}
&\mathbf{n}_1\xrightarrow{\text{cos}(\alpha)} \mathbf{P_{A}} \xrightarrow{\text{cos}(\alpha)} \mathbf{n}_1\xrightarrow{-\text{sin}(\beta)} \mathbf{P_{\bar{B}}} \xrightarrow{-\text{sin}(\beta)} \mathbf{n}_1\nonumber \\
&\mathbf{n}_2\xrightarrow{\text{sin}(\alpha)} \mathbf{P_{A}} \xrightarrow{\text{sin}(\alpha)} \mathbf{n}_2\xrightarrow{\text{cos}(\beta)} \mathbf{P_{\bar{B}}} \xrightarrow{\text{cos}(\beta)} \mathbf{n}_2\nonumber \\
&\mathbf{n}_1\xrightarrow{\text{cos}(\alpha)} \mathbf{P_{A}} \xrightarrow{\text{sin}(\alpha)} \mathbf{n}_2\xrightarrow{\text{cos}(\beta)} \mathbf{P_{\bar{B}}} \xrightarrow{-\text{sin}(\beta)} \mathbf{n}_1\nonumber \\
&\mathbf{n}_2\xrightarrow{\text{sin}(\alpha)} \mathbf{P_{A}} \xrightarrow{\text{cos}(\alpha)} \mathbf{n}_1\xrightarrow{-\text{sin}(\beta)} \mathbf{P_{\bar{B}}} \xrightarrow{\text{cos}(\beta)} \mathbf{n}_2,
\end{align}
leading to an intensity:
\begin{equation}
I_{A\bar{B}}(\tau)=|\mathbf{P}_{\bar{B}}(\tau)|^2 (-{n}_1\text{cos}(\alpha)\text{sin}(\beta) +{n}_2\text{sin}(\alpha)\text{cos}(\beta))^2.
\end{equation}

In an analogous way, the following intensities are produced in the case that photon $A$ is deflected:
\begin{align}
& I_{\bar{A}B}(\tau)=|\mathbf{P}_B(\tau)|^2(-{n}_1\text{sin}(\alpha)\text{cos}(\beta) +{n}_2\text{cos}(\alpha)\text{sin}(\beta))^2 \nonumber\\
&I_{\bar{A}\bar{B}}(\tau)=|\mathbf{P}_{\bar{B}}(\tau)|^2({n}_1\text{sin}(\alpha)\text{sin}(\beta) +{n}_2\text{cos}(\alpha)\text{cos}(\beta))^2.
\end{align}
The evolution of the polarizer state vector $\mathbf{S}_B(\tau)$ is governed by: 
\begin{equation}\label{EPRevolutionlaw2B}
    \frac{\partial \textbf{S}_{B}(\tau)}{\partial \tau} = \kappa_{meas} s_{B}(\tau) (\textbf{S}_{B}(\tau) \cdot \textbf{1}_{B})(\textbf{1}_{B} - (\textbf{S}_{B}(\tau) \cdot \textbf{1}_{B})\textbf{S}_{B}(\tau)).
\end{equation}
Similar as above, Eq.~\eqref{EPRevolutionlaw2B} can be reformulated in terms of the angle $\phi_B(\tau)$ of the vector $\mathbf{S}_B(\tau)$ with respect to the $\mathbf{h}$-axis:
\begin{equation}\label{mmtconditionphiB}
    \frac{\partial \phi_B(\tau)}{\partial \tau} = \frac{1}{2}k_{meas} s_B(\tau) \text{sin}(2(\beta-\phi_B(\tau))).
\end{equation}
At the onset of the interaction, at $\tau=\tau_B$, polarizer $P_B$ is in an undecided state, characterized by $\mathbf{P}_B(\tau_B)=\frac{1}{\sqrt{2}}\mathbf{1}_B$ and $\mathbf{P}_{\bar{B}}(\tau_B)=\frac{1}{\sqrt{2}}\mathbf{1}_{\bar{B}}$
such that $\mathbf{S}_B(\tau_B)=\frac{1}{\sqrt{2}}(\mathbf{1}_{{B}}+ \mathbf{1}_{\bar{B}})$.
However, similar as for photon $A$, Eq.~\eqref{mmtconditionphiB} forces the vector $\mathbf{S}_B(\tau)$ to align parallel with the polarizer vector $\mathbf{1}_B$, corresponding to a pass, or orthogonal to $\mathbf{1}_B$ (i.e., parallel to $\mathbf{1}_{\bar{B}}$), representing a deflection. Again, there are four possible scenarios, depending on the sign of $s_B(\tau_{B})$ and on the angle between $\mathbf{S}_B(\tau_B)$ and $\mathbf{1}_B$. If $s_B(\tau_B)$ is positive, then $\mathbf{S}_B(\tau)$ is attracted towards $\textbf{1}_B$ if $\textbf{S}_B(\tau_B)\cdot \textbf{1}_B\geq 0$ (corresponding to $|\phi_{B}(\tau_B)-\beta| \leq \pi/2$), and towards $-\mathbf{1}_B$ if $\textbf{S}_B(\tau_B)\cdot \textbf{1}_B < 0$ (corresponding to $|\phi_{B}(\tau_B)-\beta| > \pi/2$). This outcome corresponds to photon $B$ passing through the beam splitter. However, if $s_B(\tau_B)$ is negative, $\mathbf{S}_B(\tau)$ is drawn towards $\mathbf{1_{\bar{B}}}$ if $\mathbf{S}(\tau_B)\cdot \mathbf{1_{\bar{B}}} \geq 0$ and towards $-\mathbf{1_{\bar{B}}}$ if $\mathbf{S}_B(\tau_B)\cdot \mathbf{1}_{\bar{B}}<0$, which corresponds to worldline $B$ being deflected. As the orientation of $\mathbf{S}_B(\tau)$ changes as a function of $\tau$, the magnitudes of the vectors $\mathbf{P}_B(\tau)$ and $\mathbf{P}_{\bar{B}}$ change as:
\begin{align}
& \mathbf{P}_B(\tau)=(\mathbf{S}_B(\tau) \cdot \mathbf{1}_B) \mathbf{1}_B \nonumber\\
&\mathbf{P}_{\bar{B}}(\tau)=(\mathbf{S}_{B}(\tau) \cdot \mathbf{1}_{\bar{B}}) \mathbf{1}_{\bar{B}}.
\end{align}

The outcome of photon $B$ (a pass or a deflection) is determined by $s_B(\tau)=\text{sgn}(I_{AB}(\tau)-r_B|\mathbf{P}_B(\tau)|^2)$ with hidden variable $r_B$, which is a random number between 0 and 1. In the case that photon $A$ passes we find, using $s_B(\tau_B)=\text{sgn}(I_{AB}(\tau_B)-r_B|\mathbf{P}_B(\tau_B)|^2)$, that the probability for $s_B(\tau)\geq 0$ is:
\begin{equation}\label{outcomeABpass}
    p(A,B)=({n}_1\text{cos}(\alpha)\text{cos}(\beta) +{n}_2\text{sin}(\alpha)\text{sin}(\beta))^2,
    \end{equation}
while the probability for $s_B(\tau_B) < 0$ is:
    \begin{equation}\label{outcomeABdeflect}
    p(A,\bar{B})=({n}_1\text{sin}(\alpha)\text{sin}(\beta) +{n}_2\text{cos}(\alpha)\text{cos}(\beta))^2.
\end{equation}
Since the sign of $s_B(\tau_B)$ determines the outcome, Eqs.~\eqref{outcomeABpass} and \eqref{outcomeABdeflect} are the probabilities $p(A,B)$ and $p(A,\bar{B})$ for the outcomes of photon $B$ given that photon $A$ passes. 
Similarly, if photon $A$ is deflected, the probabilities for the outcomes of photon $B$ are:
\begin{align}\label{outcomesAnot}
 & p(\bar{A},B)=(-{n}_1\text{sin}(\alpha)\text{cos}(\beta) +{n}_2\text{cos}(\alpha)\text{sin}(\beta))^2 \nonumber\\
&p(\bar{A},\bar{B})=({n}_1\text{sin}(\alpha)\text{sin}(\beta) +{n}_2\text{cos}(\alpha)\text{cos}(\beta))^2.
\end{align}

The correlations between the outcomes of photons $A$ and $B$ in Eqs.~\eqref{outcomeABpass}-\eqref{outcomesAnot}, measured in spacelike separated regions, are nonlocal from the perspective of standard spacetime, but emerge in a fundamentally local way within the crystallizing spacetime framework.

\subsection{Fundamentally local action along worldlines}\label{FundLocal}

In the main text, two theoretical models replicating quantum phenomena are developed within the crystallizing spacetime framework which rely on fundamentally local interactions mediated along worldlines as a function of $\tau$. A detailed description of such fundamentally local interactions goes beyond the scope of this work. Here, it suffices to recognize that fundamentally local interactions along worldlines are possible and that these may occur in arbitrarily short intervals $\Delta \tau$ of the evolution parameter $\tau$.

One way to demonstrate this possibility is to parametrize each contributing worldline with a suitable affine parameter $\lambda$ for each value of $\tau$. Then, bits of information available at one end of the worldline, e.g. at $\lambda=0$, can be transported across the worldline as a function of $\tau$ according to $d\lambda_I/d\tau=f(\lambda_I)$, where $f(\lambda_I)$ may for example be a constant. In this way, arbitrary amounts of information can be transported across a worldline in a fundamentally local way. And, by properly choosing the function $f(\lambda)$, this transfer of information may be realized in an arbitrarily short interval $\Delta \tau$. However, also more specific mechanisms can be imagined to transport scalar or vectorial information across worldlines. 
 
In the EPR model, the aim is to reproduce nonlocal correlations without requiring instantaneous phenomena or mysterious action at a distance. Above, the simplifying assumption was made that the excitation intensity $I(\tau)$ adapts instantaneously as a function of $\tau$ to changing coupling coefficients in distant regions of spacetime. Next, it is elaborated how this excitation efficiency can be understood in a fundamentally classical way and within a finite $\tau$-interval.

\begin{figure*}[ht!]
\centering
\includegraphics[width=16.0cm]{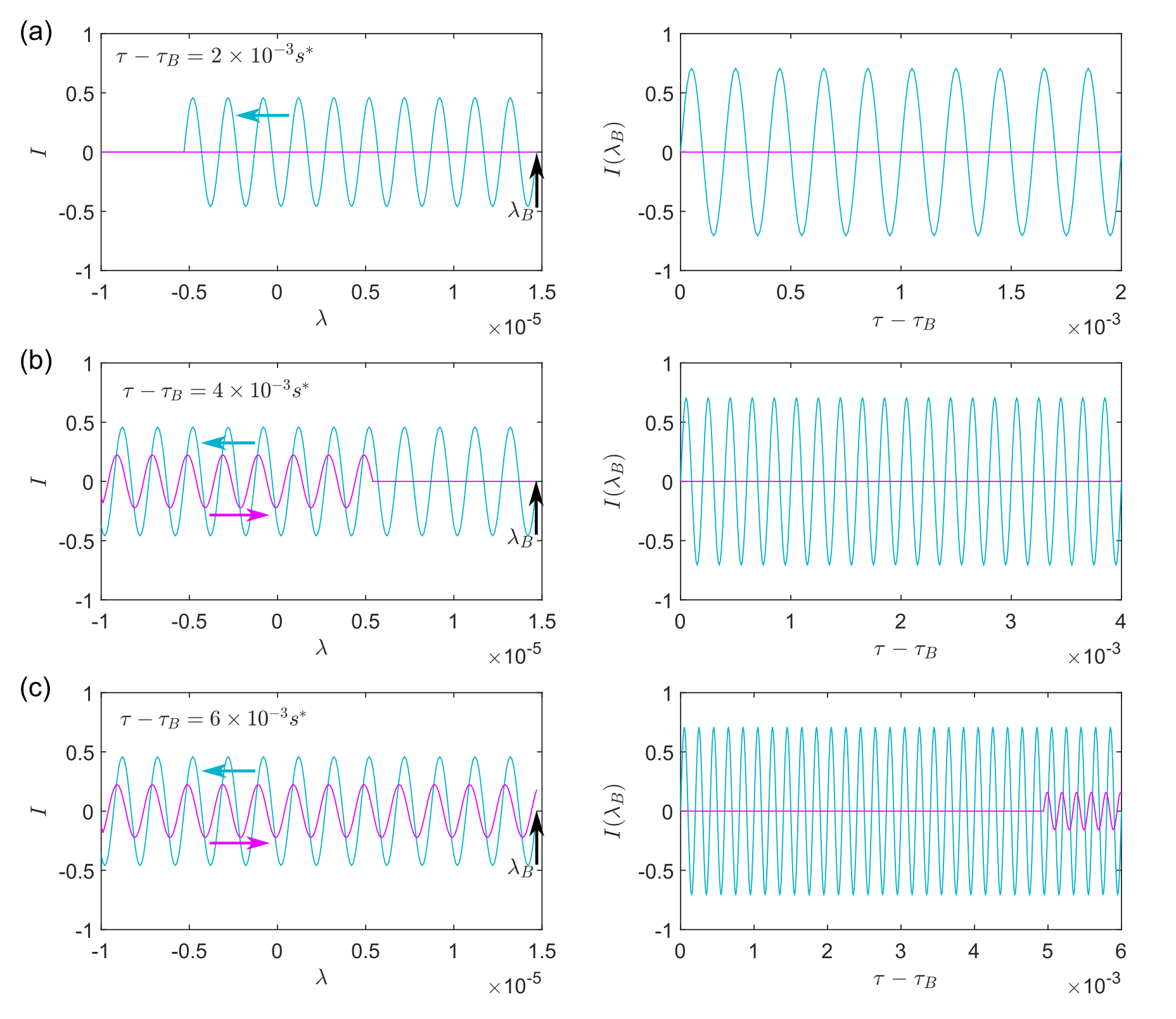}
\caption{Illustration of fundamentally classical transfer of information as a function of $\tau$ along worldlines. An affine parameter $\lambda$ is chosen between polarizer $P_A$ ($\lambda=-10^{-5}$), the source ($\lambda=0$) and polarizer $P_B$. A wave (cyan) is excited at $P_B$ and travels towards polarizer $P_A$. There a wave is produced that travels back to polarizer $P_B$. Snapshots are shown at a) $\tau=2\times 10^{-3}s^*$, b) $\tau=4\times 10^{-3}s^*$, and c) $\tau=6\times 10^{-3}s^*$ after the onset of the interaction with polarizer $P_B$. After a finite interval $\Delta \tau=5\times 10^{-3}$ the feedback wave returns with amplitude representing the excitation intensity $I$. }\label{FigSI3}
\end{figure*}

Since we operate in the fast spacetime and geodesic relaxation regime, we can consider fixed worldlines as a function of $\tau$ labeled with an affine parameter $\lambda$. Fig.~\ref{FigSI3} illustrates the worldline configuration for the case of the quantum state in Eq.~\eqref{singletgeneral}, right after the onset of the interaction with polarizer $P_B$ at $\tau=\tau_B$. We focus on just a single excitation loop, involving polarizer vector $\mathbf{P}_B$ at position $\lambda_B$ associated to a pass state and photon polarization vector $\mathbf{n}_1$. A standard wave equation is considered:
\begin{equation}\label{wave}
    \frac{\partial^2 A(\lambda,\tau)}{\partial \tau^2}=v^2\frac{\partial^2 A(\lambda,\tau)}{\partial \lambda^2},
\end{equation}
where $\lambda$ is a suitably chosen affine parameter along the loop and $v$ is a chosen wave velocity. Eq.~\eqref{wave} describes a scalar property $A(\lambda,\tau)$ along the worldline. In analogy to a standard wave, we propose a solution of the form $A(\lambda,\tau)=A_0\text{cos}(k^*\lambda-\omega^*\tau+\phi)$, where $\omega^{*2}=v^2k^{*2}$. The parameter $k^*$ may depend on the position $\lambda$ along the worldline, such that this corresponds to the local wavelength $\lambda_{ph}$ and frequency $\nu_{ph}$, whereas the remaining parameter $\omega^*$ can be chosen freely. This means that the wave velocity $v$ can be chosen arbitrarily large. In the simulation of Fig.~\ref{FigSI3} we choose a very large wavelength, for illustrative purposes.

At the onset of the interaction, i.e., at $\tau=\tau_B$, the amplitude of the excitation is $|\mathbf{P}_B|=1/\sqrt{2}$. Coupling to photon polarization vector $\mathbf{n}_1$ produces an amplitude $n_1\text{cos}(\beta)/\sqrt{2}$ of the excited wave at $\lambda_B$. The resulting wave (cyan) travels along photon worldline $B$, couples with unit efficiency to photon worldline $A$, and continues towards polarizer $P_A$ at $\lambda_A$. There, coupling to polarizer vector $\mathbf{P}_A$, which has fully collapsed to the pass state and has amplitude 1, is achieved with efficiency $\text{cos}(\alpha)$. Coupling back via the photon polarization vector $\mathbf{n}_1$ occurs with efficiency $n_1\text{cos}(\alpha)$. Consequently, a wave (magenta) with reduced amplitude travels back, via the source, towards polarizer $P_B$. Finally, coupling to the original vector $\mathbf{P}_B$ occurs with efficiency $\text{cos}(\beta)/\sqrt{2}$. The amplitude of the feedback wave is then $I=n_1^2\text{cos}^2(\alpha)\text{cos}^2(\beta)/2$. In Fig.~\ref{FigSI3}c, with the chosen value for the wave velocity $v$, the wave is found to return at $\lambda_B$ after a finite interval $\Delta \tau=5\times 10^{-3}s^*$. A similar analysis can be performed for any other excitation loop.

This demonstrates how the feedback intensity $I$, which contains information regarding the outcome at Alice's side, is available at polarizer $P_B$ after a finite $\tau$-interval. Since there is no physical restriction for how large the wave velocity can be chosen, this interval $\Delta\tau$ can be made arbitrarily short. Hence, one can make a simplifying assumption that this feedback occurs instantaneously as a function of $\tau$. 

A similar analysis of fundamentally local processes can be made in the case of the double-slit model. Here, we can assume that the intensities $I_k(\xi_k)$ are established in a fundamentally local way, since all the required phases $\phi_{ik}$ can be considered to be known scalar properties at the endpoints of worldlines $W_{ik}$. But the scalar information regarding the excitation intensities $I_{k'}\Delta\xi_{k'}$, which is available at the endpoints of the worldlines $W_{mk'}$, must still be transferred to the start points $x_m$ of the worldlines $W_{mk'}$. Such transfer of scalar information may occur in a variety of ways. For example, in analogy to well-known transport phenomena in standard spacetime like electrical current flowing through a conducting wire or liquid flowing through a pipe, one may consider the transport of a scalar property along a worldline using similar equations. Also the wave equation in Eq.~\eqref{wave} can be used for transporting a scalar property, namely the amplitude of the wave, across a worldline.

This qualitative assessment should be sufficient to demonstrate that the proposed models are fundamentally local.

\subsection{Non-superdeterminism}\label{Nonsuper}

It is important to stress that there are different ways in which statistical independence, i.e. the absence of statistical relations between measurement settings and the system's state, may be violated in Bell-type experiments.

By invoking superdeterminism one can explain the emergence of quantum correlations in Bell-type experiments, even in a classical setting of 4D spacetime~\cite{Hooft2014,Hossenfelder2020}. This is possible since superdeterminism relaxes the assumption of statistical independence of Bell's theorem, while preserving local realism. Even though different definitions of superdeterminism are sometimes used in other works, here it means that there are correlations between measurement settings and measurement outcomes, determined by pre-existing conditions, for example at the Big Bang. Even choices of measurements settings by distant operators and potential hidden variables determining the outcome of an experiment are then predetermined, precisely such that observed correlations obey the statistics of quantum theory. Even though there are no hard arguments against this possibility, in this work we will discard such superdeterminism because its conspiratorial nature is simply too counterintuitive. 

The presented crystallizing spacetime framework constitutes a more plausible fundamentally deterministic theory, in which a violation of statistical independence is realized by relying on future-input-dependence and zigzag action along worldlines. Here, the observed quantum correlations emerge as a result of physical mechanisms, occurring across spacetime as a function of $\tau$, that do not rely on predetermined settings or hidden variables. Therefore, the experimenters can freely choose their measurement settings, avoiding large conflicts with the concept of free will.

\bibliographystyle{unsrt}
\bibliography{CST}

\end{document}